\documentclass[a4paper,fleqn,usenatbib]{mnras}

\usepackage{mathptmx}

\usepackage[T1]{fontenc}
\usepackage{ae,aecompl}

\usepackage{graphicx}	
\usepackage{amsmath}	
\usepackage{amssymb}	
\usepackage{ifxetex}
\usepackage{multicol}

\def\S_AB{S$\underline{\rm A}$B}
\def\SA_B{SA$\underline{\rm B}$}
\def\SB_a{SB$_a$}

\def\^+{$^+$}
\def\^o{$^o$}
\def\_rs{$\underline{\rm r}$s}

\def\ro4r{$r_{O4R}$}
\def\ri4r{$r_{I4R}$}
\def\a_b{a$\underline{\rm b}$}

\def\apjs{{ApJS}}

\def\s4g{{S$^4$G}}

\title[Supernova 2017eaw in NGC 6946]{$BVRI$ Photometry of the Classic Type II-P
Supernova 2017eaw in NGC 6946: Day 3 to Day 594}

\author[Ronald J. Buta and William C. Keel]{
Ronald J. Buta\thanks{E-mail: rbuta@ua.edu} and William C. Keel
\\
Department of Physics \& Astronomy,University of Alabama, Box
870324, Tuscaloosa, AL 35487
}
\date{Accepted XXX. Received YYY; in original form ZZZ}

\pubyear{2016}

\begin{document}
\label{firstpage}
\pagerange{\pageref{firstpage}--\pageref{lastpage}}
\maketitle

\begin{abstract}
Broadband $BVRI$ light curves of SN 2017eaw in NGC 6946 reveal the
classic elements of a Type II-P supernova. The observations were begun
on 16 May 2017 (UT), approximately 1 day after the discovery was
announced, and the photometric monitoring was carried out over a period
of nearly 600 days. The light curves show a well-defined plateau and an
exponential tail which curves slightly at later times. An approximation
to the bolometric light curve is derived and used to estimate the
amount of $^{56}$Ni created in the explosion; from various approaches
described in the literature, we obtain $M$($^{56}{\rm Ni}$) =
0.115\rlap{$_{-{0.022}}$}$^{+{0.027}}$$M_{\odot}$.  We also estimate
that 43\% of the bolometric flux emitted during the plateau phase is
actually produced by the $^{56}$Ni chain. Other derived parameters
support the idea that the progenitor was a red supergiant.
\end{abstract}




\section{Introduction}

Type II-P supernovae are believed to be massive stars which undergo
violent core collapse at a point in their evolution when they are red
supergiants. Typically recognized by the presence of Balmer emission
lines, and showing a characteristic ``plateau" (or extended period of
nearly constant brightness - the ``P" in Type II-P) in their light
curves, Type II-P supernovae constitute about 50\% of all core-collapse
supernovae.  The ``standard model" of such supernovae is that of a
massive star that gradually builds up an iron core whose mass at some
point exceeds the Chandrasekhar limit, and collapses violently into a
neutron star. The shock wave from core collapse propagates through a
deep envelope of hydrogen that is carried into space, leaving the
neutron star as the remnant. Although models have suggested that stars
in the range 8-30$M_{\odot}$ are the likely progenitors of Type II-P
supernovae, most of the real progenitors that have been identified,
usually in {\it Hubble Space Telescope} images, are in the range
8-16M$_{\odot}$. For an excellent review of the properties of all
known types of supernovae, see Branch and Wheeler (2017).

Supernova 2017eaw was reported by Wiggins (2017) and confirmed by Dong
\& Stanek (2017) as the 10th supernova discovered in the ``Fireworks
Galaxy" NGC 6946. Wiggins' discovery image, obtained on 2017 May
14.2383 (UT), revealed a new star 153$^{\prime\prime}$ northwest of the
center of the galaxy at a magnitude of 12.8. Wiggins also obtained an
image on May 12 that revealed nothing at the position of the new star,
which pins down the explosion date to May 13$\pm$1 (UT), or JD
2457886.5 (see also Rui et al. 2019; Szalai et al. 2019; Van Dyk et
al. 2019). An optical spectrum obtained by Tomasella et al. (2017) on
2017 May 15.13095 showed a P-Cygni type H$\alpha$ profile, indicating
the supernova to be of Type II.

On May 16.256, we began a long-term campaign to monitor the $BVRI$
light curves of SN 2017eaw with the University of Alabama 0.4m DFM
Engineering Ritchey-Chretien reflector. The observatory is located on
campus atop Gallalee Hall, a severely light-polluted site.  Later
observations were made with larger telescopes at remote observatories in
Arizona and the Canary Islands.

NGC 6946, type SAB(rs)cd (de Vaucouleurs et al. 1991; Buta et al.
2007), is a prototypical late-type spiral with a large population of
massive stars. The galaxy is nearly face-on but lies at a Galactic
latitude of only 11\rlap{.}$^o$7, and hence suffers considerable foreground
extinction. The high northern declination facilitated nearly continuous
coverage of the light curves except for the month of February,
when the supernova could not be observed at night for a few weeks.

Here we present the results of our monitoring of SN 2017eaw over a
period of nearly 600 days. Tsvetkov et al. (2018) also present $UBVRI$
photometry of SN 2017eaw to 206 days, with which we find good
agreement. Other photometric observations are presented by Szalai et
al. (2019), Rui et al. (2019), and Van Dyk et al. (2019). The
observations reveal the classic light and colour evolution curves of a
typical Type II-P supernova. From our light curves we also derive
several basic parameters - e.g., absolute magnitude at maximum
brightness, slope of the post-plateau decline in brightness, mass of
$^{56}$Ni produced, explosion energy, and estimated progenitor radius -
to evaluate how SN 2017eaw compares to other supernovae of the same
type.

\section{Observations}

In addition to the UA 0.4m telescope, two telescopes operated by the
Southeastern Association for Research in Astronomy (SARA) were used for
the observations of SN 2017eaw: the SARA-KP 0.96-m telescope at Kitt
Peak, Arizona, and the SARA-RM (formerly Jacobus Kapteyn) 1.0-m
telescope at the Roque de los Muchachos on La Palma, the Canary
Islands. The SARA facilities are described by Keel et al. (2017) and
are used remotely. Table~\ref{tab:tels} summarizes the CCDs, pixel
scales, and other characteristics of the instruments used and of the
images obtained.

\begin{table*}
\centering
\caption{Properties of instruments and images. Col. 1: CCD camera;
col.  2: observatory (KP=Kitt Peak, Arizona, USA; RM=Roque de los
Muchachos, La Palma, Spain; UA = University of Alabama, Tuscaloosa,
USA); col. 3: scale in arcsec pix$^{-1}$; col. 4: field dimensions in
arcminutes covered by CCD frames; cols. 5 and 6:  CCD properties; cols.
7 and 8: mean and standard deviation of the full width at half maximum
of the point spread function on images, in pixels; and col. 9: number
of nights in the mean.
}
\label{tab:tels}
\begin{tabular}{llcccrccr}
\hline
 CCD            & Telescope       & Pixel    & Image                                              & Gain           & Read noise    & $<$FWHM$>$  & Std. Dev.     & Number \\
               &                 & scale    & dimensions                                          & ($e^-$/ADU)    & ($e^-$)  & (pix)       & (pix) & nights              \\
1 & 2 & 3 & 4 & 5 & 6 & 7 & 8 & 9 \\
\hline
ARC            &   SARA-KP 0.96-m &   0\rlap{.}$^{\prime\prime}$44  & 15\rlap{.}$^{\prime}$0$\times$15\rlap{.}$^{\prime}$0 &   2.3         &  6.0      &  7.0   &    2.3     &     6 \\
Andor Icon-L   &   SARA-RM 1.0-m &   0\rlap{.}$^{\prime\prime}$34   & 11\rlap{.}$^{\prime}$6$\times$11\rlap{.}$^{\prime}$6 &   1.0         &  6.3      &  4.6   &    1.3     &    13 \\
SBIG STL-6303E &   UA 0.4m       &   0\rlap{.}$^{\prime\prime}$57   & 29\rlap{.}$^{\prime}$2$\times$19\rlap{.}$^{\prime}$5 &   2.3         & 13.5      &  7.0   &    1.3     &    44 \\
\hline
\end{tabular}
\end{table*}

The filters used at all three observatories were manufactured by Custom
Scientific as the Johnson/Cousins/Bessell $UBVRI$ filter set. The
transmission curves for these filters are shown at the URL
www.customscientific.com/astronomy.html. The nominal photon-weighted
effective wavelengths for a spectrum flat in $F_{\lambda}$ as
calculated from these data are $B$: 438.4 nm; $V$:  561.0 nm; $R$:
648.5 nm; and $I$: 831.3 nm. Decays in the coatings on the filters used
with the UA 0.4m telescope led to large, donut-shaped flat-field
features that were of minimal consequence in the $B$, $R$, and $I$
filters but were especially severe in the $V$ filter used up until
November 2017. Filter rotation would lead to small displacements along
one axis of the CCD that led to poor flat-fielding. The problem was
minimized by exposing the twilight flats in the direction of the galaxy
and by taking the $V$-band twilight flats last and the $V$-band
observations of NGC 6946 first.  This $V$-band filter was 
replaced with another of the same detailed characteristics, but with no
flat-fielding issues. The change had little impact on the $V$-band
transformations. For example, for 33 nights prior to 1 November 2017,
the coefficients $C_{V0}$, $C_{V1}$, and $C_{V2}$ in equation 2a
(section 3) were found on average to be 19.027$\pm$0.063,
$-$0.008$\pm$0.016, and $-$0.020$\pm$0.012, respectively. The same
coefficients after 1 November 2017 were found on average to be
19.336$\pm$0.053, 0.010$\pm$0.025, and $-$0.061$\pm$0.020 for 11
nights. The main effect of the filter change was to substantially
reduce the average rms deviation of the UA 0.4m $V$-band
transformations by a factor of 2.4, from 0.031 mag to 0.013 mag.

A typical UA 0.4m observation consisted of several 6 minute exposures,
usually with more exposures in $B$ due to lower sensitivity of the CCD
at shorter wavelengths. Typical SARA-RM and SARA-KP observations
involved 2 to 5 1-5 min exposures.  The images were flat-fielded,
registered, and combined using Image Reduction and Analysis Facility
(IRAF)\footnote{IRAF is distributed by the National Optical Astronomy
Observatories, which is operated by the Association of Universities for
Research in Astronomy, under cooperative agreement with the National
Science Foundation.} routines IMARITH, IMALIGN, and IMCOMBINE.
Photometry was performed using IRAF routine PHOT.

\section{Set-up of Local Standard stars}

\subsection{Calibrations}

\begin{figure*}
\includegraphics[width=\textwidth]{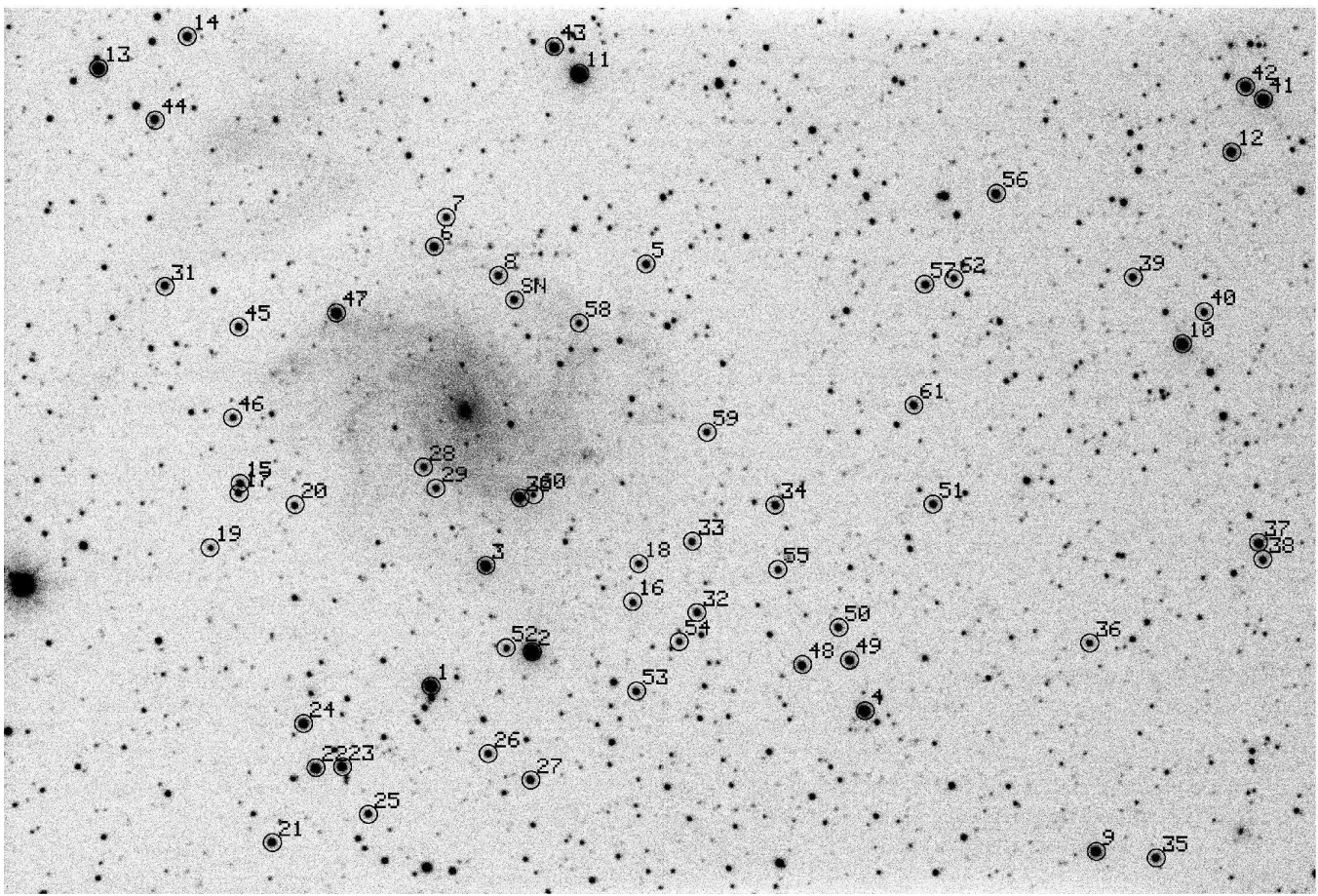}
\caption{
Local standards set up around NGC 6946. North is to the top and
east is to the left. The supernova is labeled SN on this $I$-band 
UA 0.4m image.
}
\label{fig:chart}
\end{figure*}

The light curves for SN 2017eaw are based on the secondary local
standards summarized in Table~\ref{tab:stds} and identified in
Figure~\ref{fig:chart}. These are based on averages from 5 photometric
nights for $V$, $R$, and $I$, and 4 photometric nights for $B$.
Standard stars from Landolt (1992) were used for estimating extinction
and colour terms in the transformations of these stars to the
Johnson-Cousins $BVRI$ system. The following transformation equations
were used:

$$B = b - k_Bx_B + C_{B0} + C_{B1}(B-V) + C_{B2}(B-V)^2 \eqno{1a}$$
$$V = v - k_Vx_V + C_{V0} + C_{V1}(B-V) + C_{V2}(B-V)^2 \eqno{1b}$$

$$V = v - k_Vx_V + C_{V0} + C_{V1}(V-R) + C_{V2}(V-R)^2 \eqno{2a}$$
$$R = r - k_Rx_R + C_{R0} + C_{R1}(V-R) + C_{R2}(V-R)^2 \eqno{2b}$$

$$R = r - k_Rx_R + C_{R0} + C_{R1}(R-I) + C_{R2}(R-I)^2 \eqno{3a}$$
$$I = i - k_Ix_I + C_{I0} + C_{I1}(R-I) + C_{I2}(R-I)^2 \eqno{3b}$$

\noindent
In these equations, $b$, $v$, $r$, and $i$ are the natural magnitudes
outputted by IRAF routine PHOT and reduced to an integration time of 1
sec. For the UA 0.4m, these magnitudes were derived using an aperture
of radius 15 pix with local sky readings taken in an annulus of inner
radius 20 pix and a width of 10 pix. A growth curve showed that a 15
pix radius includes 98\% of the total light of a point source on the UA
0.4m images. A similar aperture was used for the SARA-KP observations, 
but for the SARA-RM images obtained at later times we used an integration
aperture radius of 6 pix to exclude a foreground star (section 3.2.2).

The $k$ coefficients in equations 1 are atmospheric extinction
coefficients in units of magnitudes per unit airmass while the $x$
terms are the mean airmass for each filter at the midpoint of the time
of observation. The $C$s are the zero points and colour terms. The
latter are particularly important because not only does the galaxy
suffer significant foreground extinction ($A_V$ $\approx$ 1 mag), but
also the intrinsic colours of the supernova quickly reddened. This took
the supernova into the domain of the reddest Landolt standards. The
general procedure for applying equations 1-3 is to first solve for the
colour index in each pair of equations, and then derive the individual
magnitudes.

\begin{figure}
\includegraphics[width=\columnwidth]{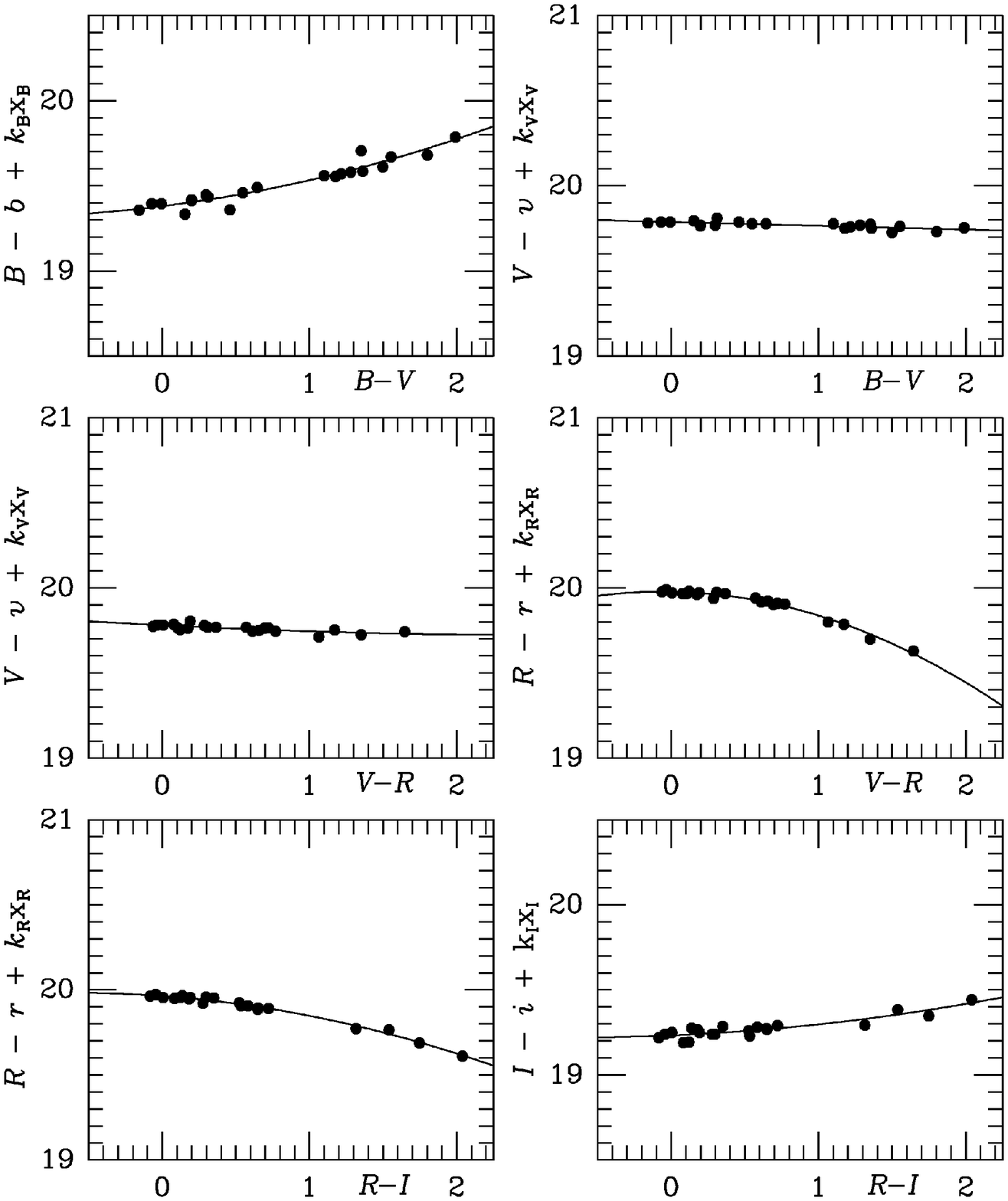}
\caption{
Graphs showing the nonlinear colour transformations 
derived from UA 0.4m observations of 15 Landolt (1992) standards
on 11 December 2017 (UT).
}
\label{fig:12-11-2017UT}
\end{figure}

The graph in Figure~\ref{fig:12-11-2017UT} shows a standard magnitude +
$kx$ (airmass term) versus standard colours. These highlight
nonlinearities in the colour terms when the reddest stars have to be
included. Table~\ref{tab:mcoeffs} summarizes mean values of the
extinction and transformation coefficients for 4 nights for $VRI$ and 3
nights for $BV$ on the UA 0.4m telescope.

\begin{table}
\centering
\caption{Mean extinction and transformation coefficients for the UA
0.4m/STL-6303E combination for the nights of 23 November 2017UT, 11
December 2017UT, 12 December 2017UT, and 13 December 2017UT.
}
\label{tab:mcoeffs}
\begin{tabular}{ccrrrcc}
\hline
Filter & $k$  & $ C_0$ & $ C_1 $ & $C_2$ & $n$ & equation \\
       & m.e. & m.e.   & m.e.    & m.e.  &     &          \\
 1     & 2    & 3      & 4       & 5     &  6  &     7    \\
\hline
$B$ &     0.300 &     19.330 &      0.092 &      0.049 &     3 & 1a\\
    &     0.027 &      0.032 &      0.007 &      0.003 &       &   \\
$V$ &     0.171 &     19.786 &   $-$0.025 &      0.000 &     3 & 1b\\
    &     0.019 &      0.019 &      0.004 &      0.000 &       &   \\
    &           &            &            &            &       &   \\  
$V$ &     0.180 &     19.793 &   $-$0.011 &   $-$0.018 &     4 & 2a\\
    &     0.015 &      0.015 &      0.027 &      0.021 &       &   \\
$R$ &     0.110 &     19.978 &   $-$0.023 &   $-$0.119 &     4 & 2b\\
    &     0.015 &      0.027 &      0.020 &      0.013 &       &   \\
    &           &            &            &            &       &   \\  
$R$ &     0.118 &     19.995 &   $-$0.075 &   $-$0.046 &     4 & 3a\\
    &     0.013 &      0.018 &      0.006 &      0.003 &       &   \\
$I$ &     0.069 &     19.235 &      0.011 &      0.031 &     4 & 3b\\
    &     0.011 &      0.011 &      0.013 &      0.002 &       &   \\
\hline
\end{tabular}
\end{table}

\begin{table*}
\centering
\caption{Local standard stars around NGC 6946. Col. 1: star number on $I$-band finding chart;
 cols. 2-5: magnitude and colours in the Johnson-Cousins $BVRI$ photometric system. The mean
error (m.e.) of each parameter is on the line below the parameter}
\label{tab:stds}
\begin{tabular}{rrcccrrcccrrccc}
\hline
No. & $V$ & $B-V$ & $V-R$ & $R-I$ & No. & $V$ & $B-V$ & $V-R$ & $R-I$ & No. & $V$ & $B-V$ & $V-R$ & $R-I$ \\
    & m.e. & m.e. & m.e.  & m.e.  &     & m.e.& m.e.  &  m.e. &  m.e. &     & m.e.& m.e.  &  m.e. &  m.e. \\
 1 & 2 & 3 & 4 & 5 &  1 & 2 & 3 & 4 & 5 &  1 & 2 & 3 & 4 & 5 \\
\hline
  1 &  10.089 &   0.219 &   0.157 &   0.160 &  22 &  12.955 &   1.824 &   1.019 &   0.895 &  43 &  11.487 &   0.660 &   0.392 &   0.314 \\
    &   0.009 &   0.009 &   0.014 &   0.011 &     &   0.010 &   0.045 &   0.010 &   0.011 &     &   0.011 &   0.007 &   0.013 &   0.017 \\
  2 &  10.432 &   1.447 &   0.784 &   0.667 &  23 &  12.676 &   1.252 &   0.711 &   0.636 &  44 &  12.666 &   0.753 &   0.481 &   0.405 \\
    &   0.013 &   0.015 &   0.007 &   0.008 &     &   0.013 &   0.022 &   0.010 &   0.011 &     &   0.016 &   0.025 &   0.011 &   0.014 \\
  3 &  11.475 &   0.593 &   0.413 &   0.359 &  24 &  12.501 &   1.093 &   0.623 &   0.548 &  45 &  13.420 &   0.906 &   0.550 &   0.502 \\
    &   0.017 &   0.007 &   0.017 &   0.012 &     &   0.012 &   0.021 &   0.011 &   0.012 &     &   0.013 &   0.017 &   0.021 &   0.016 \\
  4 &  11.775 &   1.540 &   0.869 &   0.742 &  25 &  13.341 &   0.577 &   0.368 &   0.346 &  46 &  13.795 &   0.659 &   0.412 &   0.381 \\
    &   0.012 &   0.010 &   0.012 &   0.012 &     &   0.008 &   0.012 &   0.008 &   0.016 &     &   0.011 &   0.028 &   0.010 &   0.015 \\
  5 &  13.192 &   0.728 &   0.469 &   0.409 &  26 &  13.682 &   0.738 &   0.485 &   0.439 &  47 &  13.563 &   1.924 &   1.363 &   1.608 \\
    &   0.012 &   0.015 &   0.012 &   0.016 &     &   0.018 &   0.026 &   0.017 &   0.018 &     &   0.019 &   0.039 &   0.016 &   0.015 \\
  6 &  13.117 &   1.172 &   0.677 &   0.615 &  27 &  13.231 &   0.950 &   0.585 &   0.460 &  48 &  12.491 &   0.525 &   0.347 &   0.315 \\
    &   0.013 &   0.017 &   0.015 &   0.015 &     &   0.015 &   0.019 &   0.013 &   0.015 &     &   0.007 &   0.008 &   0.010 &   0.017 \\
  7 &  13.637 &   0.767 &   0.474 &   0.434 &  28 &  13.574 &   0.721 &   0.448 &   0.417 &  49 &  13.189 &   1.454 &   0.808 &   0.699 \\
    &   0.019 &   0.008 &   0.019 &   0.013 &     &   0.015 &   0.028 &   0.010 &   0.019 &     &   0.011 &   0.018 &   0.011 &   0.015 \\
  8 &  13.272 &   0.633 &   0.441 &   0.381 &  29 &  13.774 &   0.671 &   0.456 &   0.411 &  50 &  13.646 &   1.158 &   0.794 &   0.596 \\
    &   0.015 &   0.017 &   0.017 &   0.015 &     &   0.014 &   0.030 &   0.014 &   0.016 &     &   0.016 &   0.044 &   0.018 &   0.015 \\
  9 &  11.298 &   0.290 &   0.217 &   0.226 &  30 &  13.131 &   1.802 &   1.118 &   0.989 &  51 &  13.542 &   1.035 &   0.623 &   0.560 \\
    &   0.009 &   0.013 &   0.010 &   0.014 &     &   0.013 &   0.054 &   0.012 &   0.011 &     &   0.013 &   0.024 &   0.015 &   0.014 \\
 10 &  12.653 &   1.967 &   1.164 &   1.238 &  31 &  12.960 &   0.632 &   0.443 &   0.387 &  52 &  14.000 &   0.616 &   0.421 &   0.388 \\
    &   0.012 &   0.018 &   0.017 &   0.014 &     &   0.015 &   0.012 &   0.010 &   0.015 &     &   0.015 &   0.029 &   0.012 &   0.010 \\
 11 &  10.342 &   1.470 &   0.786 &   0.691 &  32 &  13.508 &   0.869 &   0.523 &   0.483 &  53 &  13.438 &   0.632 &   0.448 &   0.390 \\
    &   0.013 &   0.010 &   0.011 &   0.012 &     &   0.016 &   0.024 &   0.012 &   0.012 &     &   0.019 &   0.009 &   0.016 &   0.013 \\
 12 &  11.823 &   0.487 &   0.354 &   0.309 &  33 &  13.488 &   0.562 &   0.386 &   0.342 &  54 &  13.909 &   0.723 &   0.426 &   0.395 \\
    &   0.014 &   0.018 &   0.016 &   0.016 &     &   0.009 &   0.008 &   0.013 &   0.017 &     &   0.018 &   0.028 &   0.020 &   0.017 \\
 13 &  10.889 &   0.817 &   0.508 &   0.401 &  34 &  12.938 &   0.635 &   0.403 &   0.355 &  55 &  13.635 &   0.532 &   0.352 &   0.338 \\
    &   0.022 &   0.017 &   0.022 &   0.013 &     &   0.009 &   0.012 &   0.010 &   0.015 &     &   0.011 &   0.007 &   0.010 &   0.016 \\
 14 &  12.277 &   0.435 &   0.323 &   0.293 &  35 &  12.653 &   0.640 &   0.411 &   0.358 &  56 &  12.763 &   0.948 &   0.609 &   0.557 \\
    &   0.025 &   0.018 &   0.024 &   0.016 &     &   0.017 &   0.010 &   0.013 &   0.017 &     &   0.008 &   0.018 &   0.012 &   0.014 \\
 15 &  13.135 &   0.655 &   0.428 &   0.382 &  36 &  13.699 &   0.686 &   0.424 &   0.386 &  57 &  13.046 &   0.700 &   0.441 &   0.386 \\
    &   0.011 &   0.012 &   0.009 &   0.015 &     &   0.019 &   0.016 &   0.020 &   0.015 &     &   0.012 &   0.019 &   0.014 &   0.016 \\
 16 &  13.804 &   0.738 &   0.464 &   0.442 &  37 &  12.893 &   1.220 &   0.800 &   0.694 &  58 &  14.124 &   0.659 &   0.406 &   0.405 \\
    &   0.011 &   0.019 &   0.014 &   0.018 &     &   0.013 &   0.007 &   0.013 &   0.012 &     &   0.011 &   0.018 &   0.016 &   0.010 \\
 17 &  14.081 &   1.333 &   0.779 &   0.679 &  38 &  13.716 &   1.198 &   0.669 &   0.609 &  59 &  13.818 &   0.742 &   0.467 &   0.403 \\
    &   0.013 &   0.038 &   0.024 &   0.016 &     &   0.020 &   0.037 &   0.022 &   0.014 &     &   0.020 &   0.019 &   0.021 &   0.014 \\
 18 &  14.366 &   1.254 &   0.699 &   0.618 &  39 &  13.700 &   1.236 &   0.706 &   0.620 &  60 &  14.826 &   0.949 &   0.587 &   0.456 \\
    &   0.018 &   0.048 &   0.009 &   0.015 &     &   0.015 &   0.005 &   0.017 &   0.013 &     &   0.013 &   0.049 &   0.025 &   0.023 \\
 19 &  14.684 &   1.074 &   0.632 &   0.591 &  40 &  13.496 &   0.696 &   0.434 &   0.401 &  61 &  13.676 &   1.264 &   0.736 &   0.639 \\
    &   0.025 &   0.047 &   0.036 &   0.018 &     &   0.016 &   0.024 &   0.014 &   0.016 &     &   0.014 &   0.030 &   0.016 &   0.016 \\
 20 &  14.262 &   1.367 &   0.716 &   0.665 &  41 &  10.938 &   0.657 &   0.435 &   0.383 &  62 &  13.688 &   0.527 &   0.390 &   0.357 \\
    &   0.020 &   0.109 &   0.028 &   0.013 &     &   0.014 &   0.018 &   0.015 &   0.015 &     &   0.011 &   0.024 &   0.018 &   0.015 \\
 21 &  12.716 &   0.596 &   0.395 &   0.360 &  42 &  11.374 &   0.516 &   0.359 &   0.321 &  .. &   ..... &   ..... &   ..... &   ..... \\
    &   0.017 &   0.012 &   0.016 &   0.014 &     &   0.013 &   0.012 &   0.016 &   0.015 &     &   ..... &   ..... &   ..... &   ..... \\
\hline
\end{tabular}
\end{table*}

The photometry in Table~\ref{tab:stds} was used exclusively to derive
the magnitudes and colours of SN 2017eaw for the UA 0.4m
observations. In these applications, it was neither necessary to know
the airmass for any filter nor was it required that a given night be
perfectly photometric. These local standards give similar results as
the primary transformations shown in Figure~\ref{fig:12-11-2017UT},
including a wide range of colour.

The Table~\ref{tab:stds} standards were also used for the SARA-RM and
SARA-KP observations.  However, many of these standards were saturated
on the images and could not be used, including the reddest stars. In
these cases, we supplemented the Table~\ref{tab:stds} standards with
fainter local standards from Pozzo et al. (2006) and Botticella et al.
(2009). In the Appendix a comparison is made between the magnitudes
and colours from these and other sources with the system in
Table~\ref{tab:stds}.  Only linear colour terms were used for the SARA
observations, especially for the SARA-RM observations, most of which
were made when the supernova was getting bluer.

\subsection{Photometry}

\subsubsection{Transformation Issues}

Standard stars like those of Landolt (1992) work well for establishing
a set of local standards around any galaxy, but this does not mean they
will work well for a supernova. The classification of any supernova as
``Type II" is based on spectroscopy, in particular the presence of
H$\alpha$ in emission. Prominent emission lines and other spectral
features of supernovae can affect the reliability of the standard star
transformations (e.g., Suntzeff et al. 1988; Stritzinger et al.
2002).

The transformation equations highlighted in the previous section give
two estimates of the $R$-band magnitude: one from the $V-R$ calibration
and one from the $R-I$ calibration. These equations also give two
estimates for the $V$-band magnitude: one from the $B-V$ calibration
and one from the $V-R$ calibration. We denote these estimates as $R1$ =
$V-(V-R)$, $R2$ = $I+(R-I)$, $V1$ = $B-(B-V)$, and $V2$ = $R+(V-R)$.

\begin{figure*}
\includegraphics[width=\textwidth]{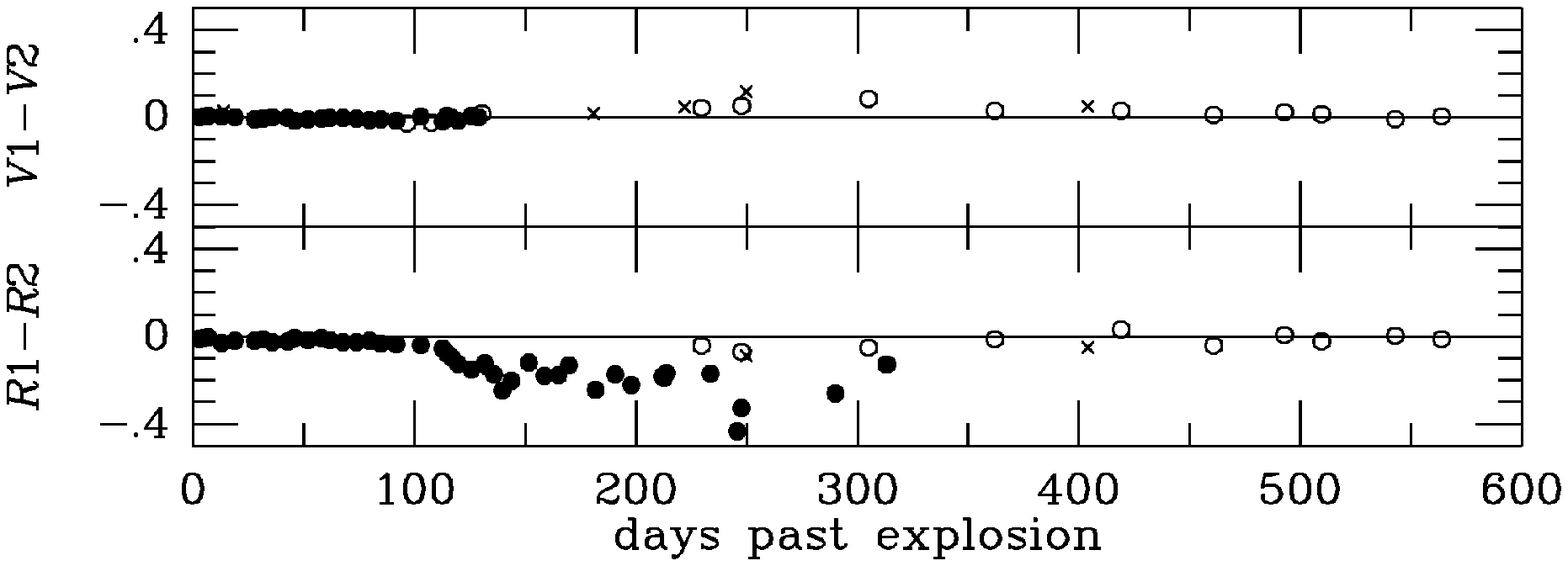}
\caption{(top) The difference between $V1=B-(B-V)$ and $V2=R+(V-R)$
and (bottom) the difference between $R1=V-(V-R)$ and $R2=I+(R-I)$, both
as a function of time after the explosion. The symbols are: UA 0.4m
observations (filled circles), SARA-KP observations (crosses), and
SARA-RM observations (open circles).
}
\label{fig:dRV12}
\end{figure*}

For normal stars, $R1$ and $R2$ (and $V1$ and $V2$) should be the same
to a few thousandths of a magnitude. The lower panel of
Figure~\ref{fig:dRV12} shows that for SN 2017eaw, the difference
$R1-R2$ is close to 0 for the first 100 days, then becomes negative a
few tenths of a magnitude from about 100-300 days, and then becomes
$\approx$0 again after 350 days. The use of the different symbols in
Figure~\ref{fig:dRV12} shows that the effect is larger for the UA 0.4m
observations compared to the SARA-KP and SARA-RM observations,
especially at 250 days past explosion. This is likely due in part to
our use of quadratic colour transformations for the UA 0.4m
observations (Figure~\ref{fig:12-11-2017UT}), compared to linear colour
transformations for the SARA observations; some of the scatter may also
be due to noise. A similar but lesser effect may be present in the
$V$-band observations (upper panel of Figure~\ref{fig:dRV12}), but
could not be followed as thoroughly as in the $R$-band because the
supernova became too faint to observe reliably in the $B$-band with the
UA 0.4m by 130 days past explosion.

The significant $R1$ and $R2$ disagreements appear just after day 100,
which corresponds approximately to the onset of the ``nebular phase" in
the supernova's evolution. At this time, the spectrum of a Type II-P
supernova (e. g., Leonard et al. 2002 and Silverman et al. 2017) begins
to show prominent emission lines of H$\alpha$, [OI] 6300, 6364, [Ca II]
7291, 7324, and the near-infrared triplet Ca II 8498, 8542, and 8662.
Szalai et al. (2019) show the spectral evolution of SN 2017eaw to 490
days past explosion, revealing that these lines did appear as expected
with H$\alpha$ and the CaII near-infrared triplet being most prominent
early-on and the [OI] and [CaII] 7291, 7324 lines appearing more
strongly later.

In view of these effects we have adopted the following procedure for
our light curves:

\noindent
1. adopt $V$ from $V-R$

\noindent
2. adopt $I$ from $R-I$

\noindent
3. adopt $B$ from $V$ and $B-V$

\noindent 
4. adopt $R1$ from $V-R$ and $R2$ from $R-I$

\noindent
5. derive $V-<R>$ and $<R>-I$ as colours, where $<R>$ is the average of
$R1$ and $R2$. The average deviation between $R1$ and $R2$ is taken
into account as part of the mean error of our $R$-band photometry.

Note that this procedure does not free our photometry from the effects
of emission lines. Instead, it makes $V-R$, $R-I$ and $V-I$ more
consistent. In general, the procedure on average reduced the values of
$V-R$ from equations 2 by factors of 0.979 before the onset of the
nebular phase, and 0.936 after onset, while the values of $R-I$ from
equations 3 were reduced by factors of 0.963 before onset and 0.860
after onset. For a few observations where only $BVR$ or $RI$ photometry
was obtained, the derived $V-R$ or $R-I$ colour index from equations 2
and 3 has been reduced slightly according to these factors, again for
consistency.

\subsubsection{Foreground Star}

\begin{figure}
\includegraphics[width=\columnwidth]{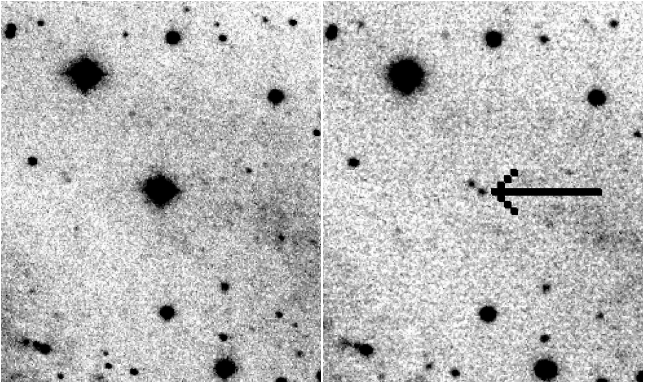}
\caption{
Comparison between SARA-RM $V$-band images of SN 2017eaw on (left) 17
August 2017 UT (day 97, $V$=13.5) and (right) 4 October 2018 UT (day
510, $V$ = 19.8), the latter revealing a foreground star ($V$ = 19.9)
close to the SN.  The field shown has dimensions 2\rlap{.}$^{\prime}$33
$\times$ 2\rlap{.}$^{\prime}$81. North is at the top and east is to the
left. The supernova is indicated by the arrow.}
\label{fig:foregrd}
\end{figure}

As SN 2017eaw faded, late-time ($>$ day 350) remote observations with
the SARA-RM telescope revealed a foreground star lying
3\rlap{.}$^{\prime\prime}$4 to the northeast
(Figure~\ref{fig:foregrd}). From six measurements, the star has a
magnitude and colours of $V$=19.90$\pm$0.04, $B-V$ = 1.04$\pm$0.04,
$V-R$ = 0.63$\pm$0.03, and $R-I$ = 0.58$\pm$0.07.  This star is close
enough to the supernova that it would have been included in the
integration apertures used for the earlier photometry.  All
observations prior to day 350 have been corrected for this star.  The
correction was made as $m_c = m_*-2.5log(x-1)$, where $m$ is the
magnitude of the supernova from equations 1-3, $m_*$ is the magnitude
of the foreground star in the same filter, and
$x$=10$^{-0.4(m-m_*)}$.  The correction generally amounted to
$\le$0.11 mag.

\section{Light Curves}

\subsection{Characteristics}

The resulting $BVRI$ photometry of SN 2017eaw is collected in
Table~\ref{tab:phot} and the light and colour curves are shown in
Figure~\ref{fig:Sgcurves5c}. The light
curves show all of the classic features of a Type II-P supernova
(Arnett 1996): a point of peak brightness followed by a rapid decline
in the $B$-band owing to cooling of the optically-thick, hot expanding
gases; a plateau, most prominent in the $R$ and $I$-bands, owing to a
drop in optical thickness as the gases cool, leading to the propogation
of a mostly hydrogen recombination front through the expanding
envelope; a sharp edge to the plateau, which occurs when the
recombination front has passed through the entire envelope
(supernebular phase); and finally, a sudden halt to the sharp decline
of the plateau edge, owing to a new important light curve power source
derived from a radioactive decay process, leading to a slow linear
decline in magnitudes called the ``tail." The plateau and the tail are
the dominant features of the light curves of Type II-P supernovae,
which represent about 40\% of all supernovae and at least 50\% of all
core collapse supernovae (Smartt et al. 2009; see also Smartt 2009; Li
et al. 2011).

The $B-V$ color evolution curve shows a fairly rapid rise from
$B-V$$\approx$0.2 at 3 days past explosion to $B-V$$\approx$2.1 at 120
days past explosion. There is no sustained period at maximum redness in
$B-V$; instead, the supernova reaches maximum $B-V$ for a few days, and
then the color declines systematically but more slowly after the onset
of the nebular phase. In contrast, both $V-R$ and $R-I$ show a
sustained period of nearly constant color index lasting from about 120
days to at least 250 days past the explosion date.

Figure~\ref{fig:tsvetkov} compares our Table~\ref{tab:phot} photometry
with the data of Tsvetkov et al. (2018), who presented $UBVRI$
observations of SN 2017eaw up to day 206. Our light curves are
well-sampled during this interval, and in Figure~\ref{fig:tsvetkov} we
compare our profiles with theirs by linearly interpolating their
profiles to the dates of our observations. The differences are plotted
as $\Delta$ = theirs (interpolated) $-$ ours, and show very good
agreement for $V$, $I$, and $V-I$. For $B$ and $R$ there are small
systematic disagreements that are likely attributable to transformation
issues. Our $R$-band magnitudes on average differ from those of
Tsvetkov et al. (2018) by 0.11$\pm$0.06 mag. This difference translates
to comparable systematic differences in $V-R$ and $R-I$. In $B-V$, the
curves are similar at early times, but by day 100 a systematic
difference of a few tenths of a magnitude appears. The SN achieves a
maximum $B-V$ in our dataset slightly redder than in the Tsvetkov et
al. dataset. A similar comparison with the photometry of Szalai et al.
(2019) gives $<\Delta$> = $-$0.05, $-$0.03, 0.06, and 0.06 mag, with
standard deviations of 0.10, 0.06, 0.07, and 0.05 mag, for $B$, $V$,
$R$, and $I$, respectively.

\subsection{Distance, Reddening, and Extinction}

The remainder of our analysis depends sensitively on the assumed
distance to NGC 6946 and the total extinction affecting SN 2017eaw.
Sahu et al. (2006) used a distance of 5.6 Mpc based on a variety of
methods for their analysis of SN 2004et. Szalai et al. (2019) adopted a
distance of 6.85$\pm$0.63 Mpc based on an average of distances
estimated from the expanding photosphere method (EPM), the standard
candle method (SCM), the tip of the red giant branch (TRGB) method, and
the planetary nebula luminosity function (PNLF). However, Eldridge and
Xiao (2019) have argued that the best distance estimate to use now for
NGC 6946 is 7.72$\pm$0.32 Mpc, based on the TRGB method (Anand et al.
2018). This is the distance we adopt in this paper.

For the reddening, we adopted $E(B-V)$ = 0.41$\pm$0.05 mag based on the
NaI D absorption line equivalent width detected towards SN 2004et
(Zwitter et al. 2004). Using Table 3 of Cardelli et al. (1989), this
translates to $A_B$ = 1.68, $A_V$ = 1.27, $A_R$= 1.06, and $A_I$ = 0.76
mag, which may be compared with the Galactic extinction values of
$E(B-V)$ = 0.30, $A_B$ =1.241, $A_V$ = 0.938, $A_R$ = 0.742, and $A_I$
= 0.515 mag, given by the NASA/IPAC Extragalactic Database (NED)
extinction calculator, based on the method of Schlafly and Finkbeiner
(2011). Although there is considerable scatter in the correlation
between NaI D equivalent width and $E(B-V)$ (Munari and Zwitter 1997),
the extreme redness of both SN 2004et and SN 2017eaw favour the higher
value of $E(B-V)$. This value was also adopted by Sahu et al. (2006)
and Szalai et al. (2019), and also is within the uncertainty of the
value of $E(B-V)$ = 0.59 $\pm$ 0.19 mag adopted by Rui et al. (2019). 

\subsection{Comparison with other SN II-P}

Figure~\ref{fig:compsnecurvesM} compares the $V$-band light curves of
SN 2017eaw with those of three other Type II-P supernovae: SN 1999em in
NGC 1637 (Leonard et al. 2002), SN2004et in NGC 6946 (Sahu et al.
2006), and SN 2012A in NGC 3239 (Tomasella et al. 2013). The magnitude
comparisons are in terms of absolute magnitudes $M_V^o$ using distance
modulus/$E(B-V)$ values of 29.44/0.41 for SN 2004et and SN 2017eaw,
29.57/0.10 for NGC 1637 and 29.96/0.037 for SN 2012A.  The results show
that the tails of SN 2004et and SN 2017eaw are very similar in $V$-band
brightness as well as shape.  On the plateau, the two supernovae differ
most significantly, by about 0.25 mag.  In contrast, SN 1999em and SN
2012A appear to have been lower luminosity Type II supernovae by nearly
2 mag.

Figure~\ref{fig:compsnecoloursM} compares the color evolution of SN
2017eaw with the same three supernovae using reddening-corrected
two-colour plots. The evolution of $(B-V)_o$ versus $(V-I)_o$
especially shows very similar curves for SN 2004et and SN 2017eaw,
except that the former is displaced blueward by 0.1-0.2 mag in
$(V-I)_o$, The points for SN 1999em and SN 2012A closely follow the
curve for SN 2017eaw. Similar results are found for $(V-R)_o$ versus
$(R-I)_o$

\subsection{Derived Parameters}

Tables~\ref{tab:params1}--~\ref{tab:nadyozhin} summarize the derived
parameters from the light curves. It appears the supernova was
discovered already very close to maximum light. Although subtle, all
four filters indicate that maximum light occurred on 2017 May 20 UT (JD
2457893.710), about 6 days after discovery.  The apparent magnitudes at
maximum light are corrected in Table~\ref{tab:params1} for the
substantial foreground extinctions.  The absolute magnitudes are close
to $-$18.0 in all four filters. This is more luminous by about 2 mag
than the average Type II SN (Li et al. 2011). Relative to maximum
light, the brightness of the SN in the plateau phase ($B_p$, $V_p$,
$R_p$, and $I_p$) is strongly dependent on passband, with $B_p -
B$(max) = 1.5 mag compared to $I_p - I$(max) = 0.05 mag.

\begin{figure}
\includegraphics[width=\columnwidth]{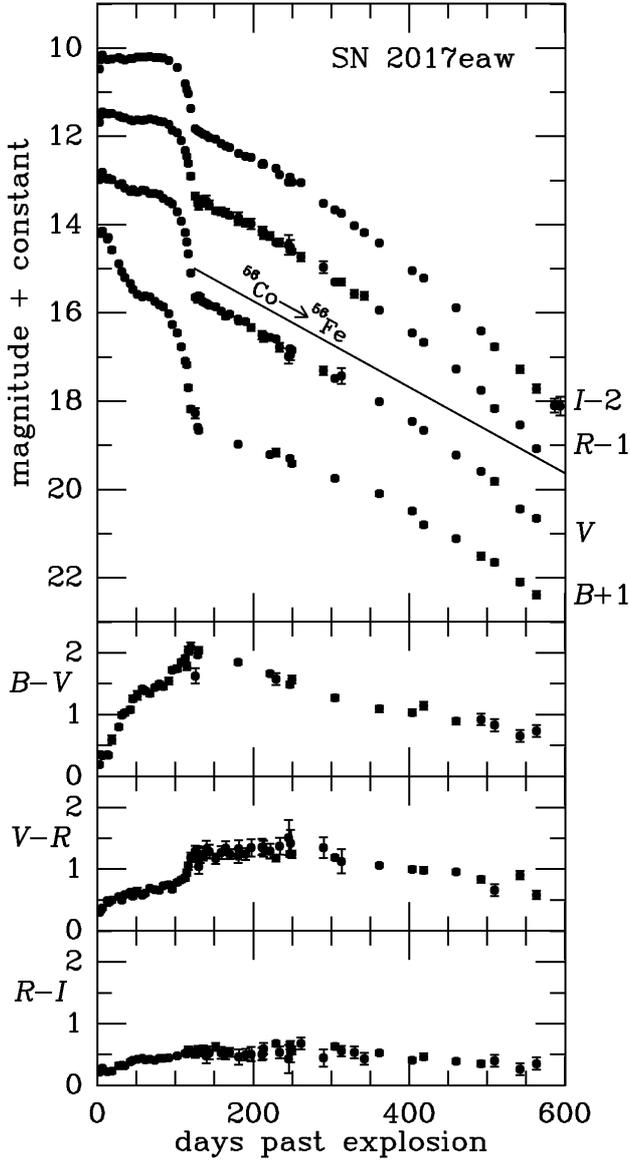}
\caption{Light and color evolution curves of SN 2017eaw to 564 days
past the explosion date for filters $B$, $V$, and $R$, and to 594 days
for filter $I$. The line shows the decline rate expected for the decay
of $^{56}$Co to $^{56}$Fe.}
\label{fig:Sgcurves5c}
\end{figure}

\begin{figure}
\includegraphics[width=\columnwidth]{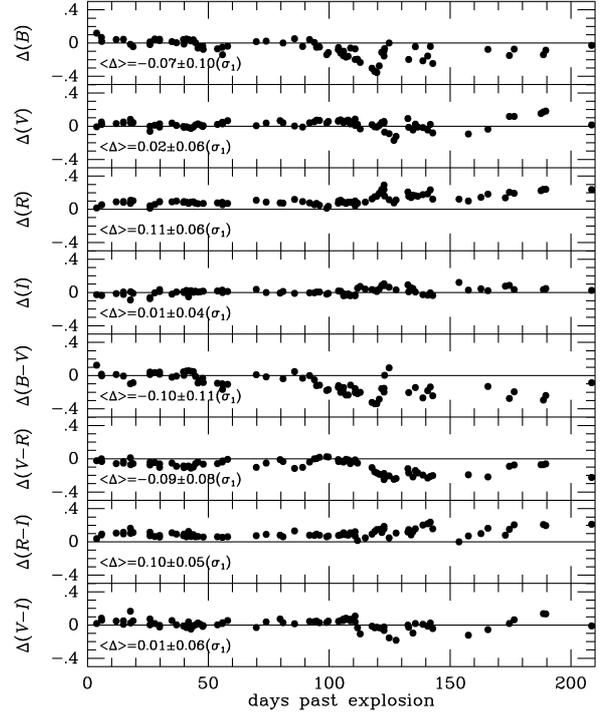}
\caption{Comparison of Table~\ref{tab:phot} photometry of SN 2017eaw
with that of Tsvetkov et al. (2018), for the first 200 days.
The average and standard deviation of each difference is shown in the lower left
of each frame.
}
\label{fig:tsvetkov}
\end{figure}

\begin{figure}
\includegraphics[width=\columnwidth]{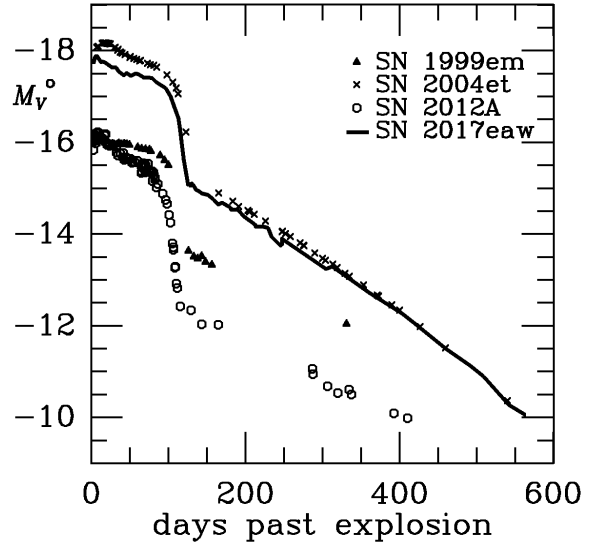}
\caption{Comparison of extinction-corrected, absolute $V$-band light
curves of SN 2017eaw with three other SN II-P.  These assume a distance
modulus of 29.44 for SN 2004et and 2017eaw (Anand et al. 2018),
29.57 for SN 1999em (Leonard et al. 2002), and 29.96 for SN 2012A
(Tomasella et al. 2013).
}
\label{fig:compsnecurvesM}
\end{figure}

\begin{figure*}
\includegraphics[width=\textwidth]{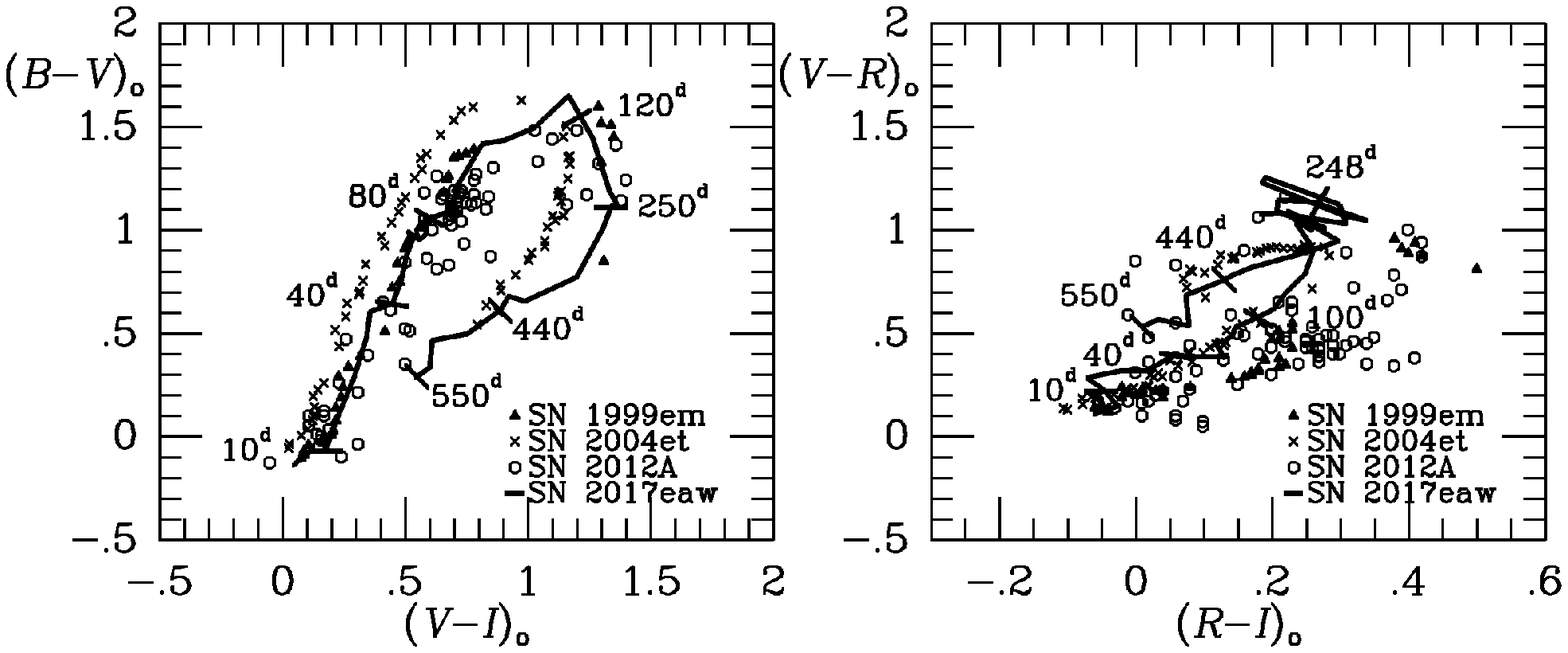}
\caption{Comparison of the extinction-corrected colour-colour evolution
of SN 2017eaw (dark solid curve) with the same evolutionary data for
three other SN II-P. The tic marks indicate specific phases, which are
labeled in days past the explosion date.
}
\label{fig:compsnecoloursM}
\end{figure*}

Past the plateau phase, the size of the dropoff relative to the plateau
level is strongly wavelength dependent, ranging from $B_r - B_p$ = 2.84
mag to $I_r-I_p$ = 1.61 mag (Table~\ref{tab:params2}). The tail begins
at $\approx$125 days past explosion and continues to the end of the
observing period, 469 days later. The light curves in this phase are
generally interpreted as being powered by the decay of radioactive
cobalt isotope $^{56}$Co into stable iron isotope $^{56}$Fe.  The
$^{56}$Co is believed to have been produced by the decay of $^{56}$Ni,
the main isotope of iron group elements explosively produced by the
post-core collapse shock wave running through the star (e.g.,
Jerkstrand 2011). The decay of $^{56}$Ni into $^{56}$Co is rapid, with
a half-life of 6 days, while the decay of $^{56}$Co into $^{56}$Fe is
77 days (Nadyozhin 1994). If the gamma rays produced by the latter
decay process are completely confined, then the light curves will show
a linear decline at a rate of 0.98 mag (100 days)$^{-1}$ (line shown in
Figure~\ref{fig:Sgcurves5c}). Since this is generally observed, the
slow, post plateau decline is often referred to as the ``radioactive
tail" of the light curves.

The properties of the tail seen in the light curves of SN 2017eaw
support this idea. The results of linear least squares fits to the tail
are summarized in Table~\ref{tab:params2}, which lists the slopes
$\gamma_B$, $\gamma_V$, $\gamma_R$, and $\gamma_I$ in units of
magnitudes per 100 days. The decline is nearly linear in the $V$ band,
but in $R$ and $I$ there appears to be a slight bend in the tail
starting about 290 days past explosion. For this reason,
Table~\ref{tab:params2} summarizes the parameters of the tail for two
parts: the earlier part from 125-290 days after the explosion, and the
later part from 290-564 days after the explosion. If the light curves
past the plateau phase are due to a radioactive decay process, then
(1.0857/slope)(ln2) gives an estimate of the half-life, $t_{1\over 2}$,
of the process. The $B$-band slope is not well-determined in the
earlier phase, but the $V$, $R$, and $I$ filters give values of
$t_{1\over 2}$ ranging from 75.8 days to 82.2 days in this phase. These
correspond to decline rates of $\gamma_V$ = 0.992 mag (100 days)$^{-1}$
and $\gamma_I$ = 0.924 mag (100 days)$^{-1}$, respectively. A combined
$VRI$ fit using the mean early tail colours as offsets gives $t_{1\over
2}$ = 79.8$\pm$1.0 days, close to the nominal value for the decay of
$^{56}$Co to $^{56}$Fe.

Past 290 days, the colours of the tail are generally bluer on average
than the colours before 290 days. The decline rates also increase;
e.g., $\gamma_V$ increases to 1.27 mag (100 days)$^{-1}$ while
$\gamma_I$ increases to 1.56 mag (100 days)$^{-1}$. This change in
decline rate could in part signify the breakdown of the gamma ray
confinement assumption, at least at later times (Woosley et al. 1989;
Sahu et al. 2006; Otsuka et al. 2012).

\begin{table*}
\centering
\caption{Photometry of SN 2017eaw. Col. 1: Universal Time date; col. 2: number of days after explosion;
cols. 3-6, line 1: magnitudes (corrected for foreground star) of SN 2017eaw; cols 3-6, line 2:
errors on magnitudes; col. 7: telescope used}
\label{tab:phot}
\begin{tabular}{ccrrrrl}
\hline
Date & Phase &  $B$ & $V$ & $R$ & $I$ & Tel \\
(UT) & JD2457886.5+ &          &     &     &     &            \\
 1 & 2 & 3 & 4 & 5 & 6 & 7 \\
\hline
05-16-2017 &     3.256 & 13.171 & 12.981 & 12.683 & 12.469 & UA 0.4m  \\
           &           &  0.024 &  0.024 &  0.017 &  0.024 & \\
05-18-2017 &     5.195 & 13.204 & 12.856 & 12.516 & 12.259 & UA 0.4m  \\
           &           &  0.025 &  0.023 &  0.022 &  0.024 & \\
05-20-2017 &     7.210 & 13.149 & 12.817 & 12.444 & 12.162 & UA 0.4m  \\
           &           &  0.031 &  0.026 &  0.014 &  0.021 & \\
05-26-2017 &    13.200 & 13.322 & 12.972 & 12.479 & 12.257 & UA 0.4m  \\
           &           &  0.020 &  0.015 &  0.017 &  0.015 & \\
05-27-2017 &    14.403 & 13.292 & 12.947 & 12.490 & ...... & SARA-KP  \\
           &           &  0.027 &  0.017 &  0.024 & ...... & \\
06-01-2017 &    19.162 & 13.575 & 12.980 & 12.480 & 12.252 & UA 0.4m  \\
           &           &  0.056 &  0.036 &  0.020 &  0.035 & \\
06-10-2017 &    28.123 & 13.888 & 13.091 & 12.534 & 12.217 & UA 0.4m  \\
           &           &  0.021 &  0.029 &  0.017 &  0.051 & \\
06-14-2017 &    32.105 & 14.068 & 13.074 & 12.571 & 12.245 & UA 0.4m  \\
           &           &  0.030 &  0.033 &  0.015 &  0.019 & \\
06-18-2017 &    36.074 & 14.201 & 13.172 & 12.587 & 12.269 & UA 0.4m  \\
           &           &  0.036 &  0.020 &  0.018 &  0.014 & \\
06-25-2017 &    43.102 & 14.331 & 13.252 & 12.628 & 12.241 & UA 0.4m  \\
           &           &  0.032 &  0.031 &  0.020 &  0.017 & \\
06-28-2017 &    46.106 & 14.466 & 13.209 & 12.645 & 12.238 & UA 0.4m  \\
           &           &  0.041 &  0.032 &  0.021 &  0.022 & \\
07-04-2017 &    52.110 & 14.578 & 13.265 & 12.631 & 12.207 & UA 0.4m  \\
           &           &  0.032 &  0.065 &  0.014 &  0.021 & \\
07-10-2017 &    58.221 & 14.634 & 13.216 & 12.640 & 12.202 & UA 0.4m  \\
           &           &  0.028 &  0.027 &  0.011 &  0.017 & \\
07-14-2017 &    62.090 & 14.617 & 13.232 & 12.621 & 12.207 & UA 0.4m  \\
           &           &  0.029 &  0.029 &  0.019 &  0.014 & \\
07-20-2017 &    68.098 & 14.636 & 13.295 & 12.607 & 12.188 & UA 0.4m  \\
           &           &  0.034 &  0.032 &  0.021 &  0.012 & \\
07-26-2017 &    74.123 & 14.733 & 13.295 & 12.625 & 12.214 & UA 0.4m  \\
           &           &  0.030 &  0.026 &  0.022 &  0.013 & \\
08-01-2017 &    80.103 & 14.816 & 13.317 & 12.658 & 12.219 & UA 0.4m  \\
           &           &  0.026 &  0.029 &  0.013 &  0.011 & \\
08-06-2017 &    85.103 & 14.855 & 13.403 & 12.667 & 12.230 & UA 0.4m  \\
           &           &  0.036 &  0.019 &  0.025 &  0.014 & \\
08-13-2017 &    92.089 & 15.016 & 13.474 & 12.726 & 12.278 & UA 0.4m  \\
           &           &  0.035 &  0.044 &  0.022 &  0.013 & \\
08-17-2017 &    96.939 & 15.258 & 13.536 & 12.858 & ...... & SARA-RM  \\
           &           &  0.042 &  0.020 &  0.024 & ...... & \\
08-24-2017 &   103.076 & 15.454 & 13.708 & 12.921 & 12.442 & UA 0.4m  \\
           &           &  0.040 &  0.029 &  0.025 &  0.012 & \\
08-28-2017 &   107.947 & 15.767 & 13.920 & 13.095 & ...... & SARA-RM  \\
           &           &  0.045 &  0.019 &  0.024 & ...... & \\
09-03-2017 &   113.069 & 16.092 & 14.185 & 13.325 & 12.808 & UA 0.4m  \\
           &           &  0.042 &  0.035 &  0.032 &  0.011 & \\
09-05-2017 &   115.089 & 16.172 & 14.392 & 13.450 & 12.938 & UA 0.4m  \\
           &           &  0.051 &  0.032 &  0.042 &  0.011 & \\
09-07-2017 &   117.094 & 16.694 & 14.660 & 13.612 & 13.036 & UA 0.4m  \\
           &           &  0.060 &  0.033 &  0.050 &  0.019 & \\
09-10-2017 &   120.082 & 17.178 & 15.093 & 13.903 & 13.364 & UA 0.4m  \\
           &           &  0.070 &  0.044 &  0.065 &  0.012 & \\
09-16-2017 &   126.098 & 17.273 & 15.646 & 14.353 & 13.826 & UA 0.4m  \\
           &           &  0.114 &  0.044 &  0.078 &  0.015 & \\
09-19-2017 &   129.113 & 17.590 & 15.625 & 14.469 & ...... & UA 0.4m  \\
           &           &  0.034 &  0.016 &  0.093 & ...... & \\
09-19-2017 &   129.205 & ...... & 15.677 & 14.428 & 13.870 & SARA-KP  \\
           &           & ...... &  0.091 &  0.077 &  0.016 & \\
\hline
\end{tabular}
\end{table*}

\begin{table*}
\centering
\setcounter{table}{3}
\caption{(cont.) Photometry of SN 2017eaw. Col. 1: Universal Time date; col. 2: number of days after explosion;
cols. 3-6, line 1: magnitudes; cols 3-6, line 2:
errors on magnitudes; col. 7: telescope used}
\label{tab:phot}
\begin{tabular}{ccrrrrl}
\hline
Date & Phase &  $B$ & $V$ & $R$ & $I$ & Tel \\
(UT) & JD2457886.5+ &  &    &     &       &            \\
 1 & 2 & 3 & 4 & 5 & 6 & 7 \\
\hline
09-20-2017 &   130.841 & 17.656 & 15.613 & 14.572 & ...... & SARA-RM  \\
           &           &  0.046 &  0.018 &  0.093 & ...... & \\
09-22-2017 &   132.172 & ...... & 15.634 & 14.454 & 13.904 & UA 0.4m  \\
           &           & ...... &  0.053 &  0.064 &  0.016 & \\
09-26-2017 &   136.169 & ...... & 15.751 & 14.513 & 13.952 & UA 0.4m  \\
           &           & ...... &  0.050 &  0.088 &  0.016 & \\
09-30-2017 &   140.057 & ...... & 15.780 & 14.459 & 13.976 & UA 0.4m  \\
           &           & ...... &  0.053 &  0.124 &  0.014 & \\
10-04-2017 &   144.169 & ...... & 15.832 & 14.552 & 14.028 & UA 0.4m  \\
           &           & ...... &  0.047 &  0.103 &  0.014 & \\
10-12-2017 &   152.076 & ...... & 15.859 & 14.690 & 14.068 & UA 0.4m  \\
           &           & ...... &  0.054 &  0.062 &  0.014 & \\
10-19-2017 &   159.061 & ...... & 15.963 & 14.698 & 14.169 & UA 0.4m  \\
           &           & ...... &  0.047 &  0.092 &  0.016 & \\
10-25-2017 &   165.168 & ...... & 16.074 & 14.737 & 14.225 & UA 0.4m  \\
           &           & ...... &  0.069 &  0.090 &  0.015 & \\
10-30-2017 &   170.034 & ...... & 16.030 & 14.786 & 14.250 & UA 0.4m  \\
           &           & ...... &  0.045 &  0.068 &  0.017 & \\
11-10-2017 &   181.090 & 17.979 & 16.136 & 14.921 & ...... & SARA-KP  \\
           &           &  0.033 &  0.018 &  0.093 & ...... & \\
11-11-2017 &   182.033 & ...... & 16.184 & 14.849 & 14.388 & UA 0.4m  \\
           &           & ...... &  0.038 &  0.123 &  0.018 & \\
11-20-2017 &   191.037 & ...... & 16.196 & 14.954 & 14.455 & UA 0.4m  \\
           &           & ...... &  0.042 &  0.089 &  0.014 & \\
11-27-2017 &   198.072 & ...... & 16.328 & 14.982 & 14.475 & UA 0.4m  \\
           &           & ...... &  0.079 &  0.115 &  0.023 & \\
12-11-2017 &   212.005 & ...... & 16.502 & 15.143 & 14.638 & UA 0.4m  \\
           &           & ...... &  0.092 &  0.095 &  0.025 & \\
12-12-2017 &   213.030 & ...... & 16.497 & 15.143 & 14.615 & UA 0.4m  \\
           &           & ...... &  0.059 &  0.098 &  0.018 & \\
12-13-2017 &   214.039 & ...... & 16.564 & 15.232 & 14.637 & UA 0.4m  \\
           &           & ...... &  0.094 &  0.089 &  0.022 & \\
12-21-2017 &   222.123 & 18.214 & 16.551 & 15.258 & ...... & SARA-KP  \\
           &           &  0.036 &  0.020 &  0.093 & ...... & \\
12-28-2017 &   229.822 & 18.162 & 16.586 & 15.405 & 14.728 & SARA-RM  \\
           &           &  0.091 &  0.037 &  0.039 &  0.018 & \\
01-02-2018 &   234.026 & ...... & 16.776 & 15.404 & 14.868 & UA 0.4m  \\
           &           & ...... &  0.101 &  0.090 &  0.023 & \\
01-14-2018 &   246.023 & ...... & 16.975 & 15.465 & 15.037 & UA 0.4m  \\
           &           & ...... &  0.170 &  0.224 &  0.028 & \\
01-15-2018 &   247.823 & 18.297 & 16.812 & 15.568 & 14.923 & SARA-RM  \\
           &           &  0.039 &  0.022 &  0.046 &  0.042 & \\
01-16-2018 &   248.026 & ...... & 16.942 & 15.519 & 14.973 & UA 0.4m  \\
           &           & ...... &  0.130 &  0.169 &  0.025 & \\
01-18-2018 &   250.095 & 18.409 & 16.842 & 15.602 & 15.034 & SARA-KP  \\
           &           &  0.056 &  0.033 &  0.054 &  0.034 & \\
01-30-2018 &   262.017 & ...... & ...... & 15.733 & 15.052 & UA 0.4m  \\
           &           & ...... & ...... &  0.093 &  0.035 & \\
02-27-2018 &   290.425 & ...... & 17.308 & 15.962 & 15.518 & UA 0.4m  \\
           &           & ...... &  0.095 &  0.136 &  0.029 & \\
03-14-2018 &   305.252 & 18.752 & 17.483 & 16.296 & 15.663 & SARA-RM  \\
           &           &  0.039 &  0.023 &  0.031 &  0.034 & \\
03-22-2018 &   313.369 & ...... & 17.431 & 16.304 & 15.746 & UA 0.4m  \\
           &           & ...... &  0.179 &  0.077 &  0.030 & \\
04-08-2018 &   330.390 & ...... & ...... & 16.567 & 16.028 & UA 0.4m  \\
           &           & ...... & ...... &  0.093 &  0.034 & \\
04-21-2018 &   343.291 & ...... & ...... & 16.614 & 16.179 & UA 0.4m  \\
           &           & ...... & ...... &  0.093 &  0.044 & \\
\hline
\end{tabular}
\end{table*}

\begin{table*}
\centering
\setcounter{table}{3}
\caption{(cont.) Photometry of SN 2017eaw. Col. 1: Universal Time date; col. 2: number of days after explosion;
cols. 3-6, line 1: magnitudes; cols 3-6, line 2:
errors on magnitudes; col. 7: telescope used}
\label{tab:phot}
\begin{tabular}{ccrrrrl}
\hline
Date & Phase &  $B$ & $V$ & $R$ & $I$ & Tel \\
(UT) & JD2457886.5+ &  &     &     &    &            \\
 1 & 2 & 3 & 4 & 5 & 6 & 7 \\
\hline
05-10-2018 &   362.199 & 19.097 & 18.005 & 16.942 & 16.418 & SARA-RM  \\
           &           &  0.047 &  0.018 &  0.022 &  0.024 & \\
06-21-2018 &   404.248 & 19.490 & 18.455 & 17.457 & 17.046 & SARA-KP  \\
           &           &  0.047 &  0.025 &  0.033 &  0.018 & \\
07-06-2018 &   419.107 & 19.800 & 18.655 & 17.672 & 17.207 & SARA-RM  \\
           &           &  0.058 &  0.025 &  0.045 &  0.028 & \\
08-17-2018 &   461.014 & 20.112 & 19.220 & 18.267 & 17.878 & SARA-RM  \\
           &           &  0.045 &  0.027 &  0.029 &  0.028 & \\
09-17-2018 &   492.887 & 20.510 & 19.588 & 18.755 & 18.404 & SARA-RM  \\
           &           &  0.082 &  0.037 &  0.033 &  0.034 & \\
10-04-2018 &   509.866 & 20.647 & 19.819 & 19.162 & 18.763 & SARA-RM  \\
           &           &  0.068 &  0.069 &  0.065 &  0.067 & \\
11-06-2018 &   542.878 & 21.095 & 20.440 & 19.538 & 19.277 & SARA-RM  \\
           &           &  0.074 &  0.057 &  0.038 &  0.087 & \\
11-27-2018 &   563.858 & 21.388 & 20.654 & 20.071 & 19.716 & SARA-RM  \\
           &           &  0.083 &  0.051 &  0.043 &  0.094 & \\
12-20-2018 &   586.846 & ...... & ...... & ...... & 20.105 & SARA-RM  \\
           &           & ...... & ...... & ...... &  0.156 & \\
12-27-2018 &   593.811 & ...... & ...... & ...... & 20.113 & SARA-RM  \\
           &           & ...... & ...... & ...... &  0.210 & \\
\hline
\end{tabular}
\end{table*}

\begin{table}
\centering
\caption{Parameters at Maximum Light
}
\label{tab:params1}
\begin{tabular}{lr}
\hline
Parameter & Value \\
1 & 2 \\
\hline
Date of explosion     & JD 2457886.5 $\pm$ 1.0 \\
Date of maximum light & JD 2457893.710 \\
Assumed distance modulus &   29.44 \\
Assumed total reddening $E(B-V)$ & 0.41$\pm$0.05 mag \\
$B$(max) &   13.15 \\
$V$(max) &   12.82 \\
$R$(max) &   12.44 \\
$I$(max) &   12.16 \\
$B_o$(max) &   11.47 \\
$V_o$(max) &   11.55 \\
$R_o$(max) &   11.38 \\
$I_o$(max) &   11.40 \\
$(B-V)_o$(max) &   $-$0.08 \\
$(V-R)_o$(max) &    0.16 \\
$(R-I)_o$(max) &   $-$0.01 \\
$(V-I)_o$(max) &    0.15 \\
$B_p - B$(max) &   1.46 \\
$V_p - V$(max) &   0.43 \\
$R_p - R$(max) &   0.18 \\
$I_p - I$(max) &   0.05 \\
$M_B^o$(max) & $-$17.97 \\
$M_V^o$(max) & $-$17.89 \\
$M_R^o$(max) & $-$18.06 \\
$M_I^o$(max) & $-$18.04 \\
\hline
\end{tabular}
\end{table}

\begin{table}
\centering
\caption{Parameters of Radioactive Tail
}
\label{tab:params2}
\begin{tabular}{lr}
\hline
Parameter & Value \\
1 & 2 \\
\hline
Onset of radioactive tail & JD 2458009.5$\pm$3          \\
Early radioactive tail &  125-291 days \\
$B_r - B_p$ &   2.84 \\
$V_r - V_p$ &   2.34 \\
$R_r - R_p$ &   1.76 \\
$I_r - I_p$ &   1.61 \\
$t_{1\over 2}$($B$) & 113.9$\pm$14.3 days   ($n$ =  8) \\
$t_{1\over 2}$($V$) & 75.8$\pm$1.6 days   ($n$ = 27) \\
$t_{1\over 2}$($R$) & 82.2$\pm$1.7 days   ($n$ = 28) \\
$t_{1\over 2}$($I$) & 81.5$\pm$1.6 days   ($n$ = 24) \\
$t_{1\over 2}$($VRI$) & 79.8$\pm$1.0 days   ($n$ = 79) \\
$<B-V>$ & 1.69$\pm$ 0.07  ($n$  =  8) \\
$<V-R>$ & 1.26$\pm$ 0.02  ($n$  = 27) \\
$<R-I>$ & 0.54$\pm$ 0.01  ($n$  = 24) \\
$<V-I>$ & 1.82$\pm$ 0.01  ($n$  = 23) \\
$<(B-V)_o>$ & 1.28$\pm$ 0.07  ($n$  =  8) \\
$<(V-R)_o>$ & 1.05$\pm$ 0.02  ($n$  = 27) \\
$<(R-I)_o>$ & 0.25$\pm$ 0.01  ($n$  = 24) \\
$<(V-I)_o>$ & 1.31$\pm$ 0.01  ($n$  = 23) \\
$\gamma_B$ & 0.661$\pm$0.082 mag (100 days)$^{-1}$ \\
$\gamma_V$ & 0.992$\pm$0.020 mag (100 days)$^{-1}$ \\
$\gamma_R$ & 0.915$\pm$0.019 mag (100 days)$^{-1}$ \\
$\gamma_I$ & 0.924$\pm$0.018 mag (100 days)$^{-1}$ \\
$\gamma_{VRI}$ & 0.943$\pm$0.012 mag (100 days)$^{-1}$ \\
 & \\
Late radioactive tail &  290-564 days \\
$<B-V>$ & 0.95$\pm$ 0.07  ($n$  =  9) \\
$<V-R>$ & 0.98$\pm$ 0.05  ($n$  = 13) \\
$<R-I>$ & 0.44$\pm$ 0.03  ($n$  = 13) \\
$<V-I>$ & 1.39$\pm$ 0.09  ($n$  = 11) \\
$<(B-V)_o>$ & 0.54$\pm$ 0.07  ($n$  =  9) \\
$<(V-R)_o>$ & 0.76$\pm$ 0.05  ($n$  = 13) \\
$<(R-I)_o>$ & 0.15$\pm$ 0.03  ($n$  = 13) \\
$<(V-I)_o>$ & 0.88$\pm$ 0.09  ($n$  = 11) \\
$\gamma_B$ & 1.071$\pm$0.032 mag (100 days)$^{-1}$ \\
$\gamma_V$ & 1.269$\pm$0.030 mag (100 days)$^{-1}$ \\
$\gamma_R$ & 1.469$\pm$0.031 mag (100 days)$^{-1}$ \\
$\gamma_I$ & 1.557$\pm$0.025 mag (100 days)$^{-1}$ \\
$\gamma_{VRI}$ & 1.427$\pm$0.039 mag (100 days)$^{-1}$ \\
\hline
\end{tabular}
\end{table}


\section{Bolometric Luminosity}
\subsection{Light Curve}

The $BVRI$ light curves of SN 2017eaw can be used to estimate its
bolometric luminosity as a function of time. This is useful for
comparing the supernova's evolution with available hydrodynamic models,
to estimate the mass of $^{56}$Ni synthesized in the explosion, and for
singular comparisons with other supernovae where the individual light
curves might be very different. Deriving a bolometric light curve for a
supernova technically requires having light curves available not only
in the optical realm, but also in the UV and IR realms. Since we only
have four filters, the construction of a bolometric light curve for SN
2017eaw will depend on bolometric corrections deduced from observations
and models of other well-observed supernovae.

Lusk and Baron (2017) describe several approaches to deriving a
bolometric light curve for a supernova. We use three approaches here.
In the first, we convert the extinction-corrected broadband magnitude
at each epoch into a flux at the top of the atmosphere using the known
flux zero points of the standard photometric systems. For $UBVRIJHKL$
(the Cousins-Glass-Johnson system), these zero points and their
effective wavelengths are given in Table A2 of Bessell et al.
(1998). Only the $R$-band in Table~\ref{tab:phot} has a measured
magnitude at every epoch of our observations. For $B$, $V$, and $I$,
linear interpolation was used to approximately fill in missing
magnitudes. The integrated luminosity from $B$ ($\lambda_{eff}$ = 0.438
nm) to $I$ ($\lambda_{eff}$ = 0.798 nm) was derived at each epoch using
the trapezoidal rule. This quasi-bolometric light curve is shown as the
light solid curve in Figure~\ref{fig:bolometricM}.

The second approach we use is a colour-bolometric correction method. In
this approach, well-observed supernovae are used to derive bolometric
corrections to a specific filter light curve as a function of colour.
These corrections are then fitted with a polynomial of relatively low
order. Lyman et al. (2014) were able to derive a fairly well-defined
parabolic relation between the $B$-band bolometric correction of Type
II SNe and the extinction-corrected $(B-I)_o$ colour index,

$${\rm BC}_B = 0.004 - 0.297(B-I)_o - 0.149(B-I)_o^2,\eqno{4}$$

\begin{figure}
\includegraphics[width=\columnwidth]{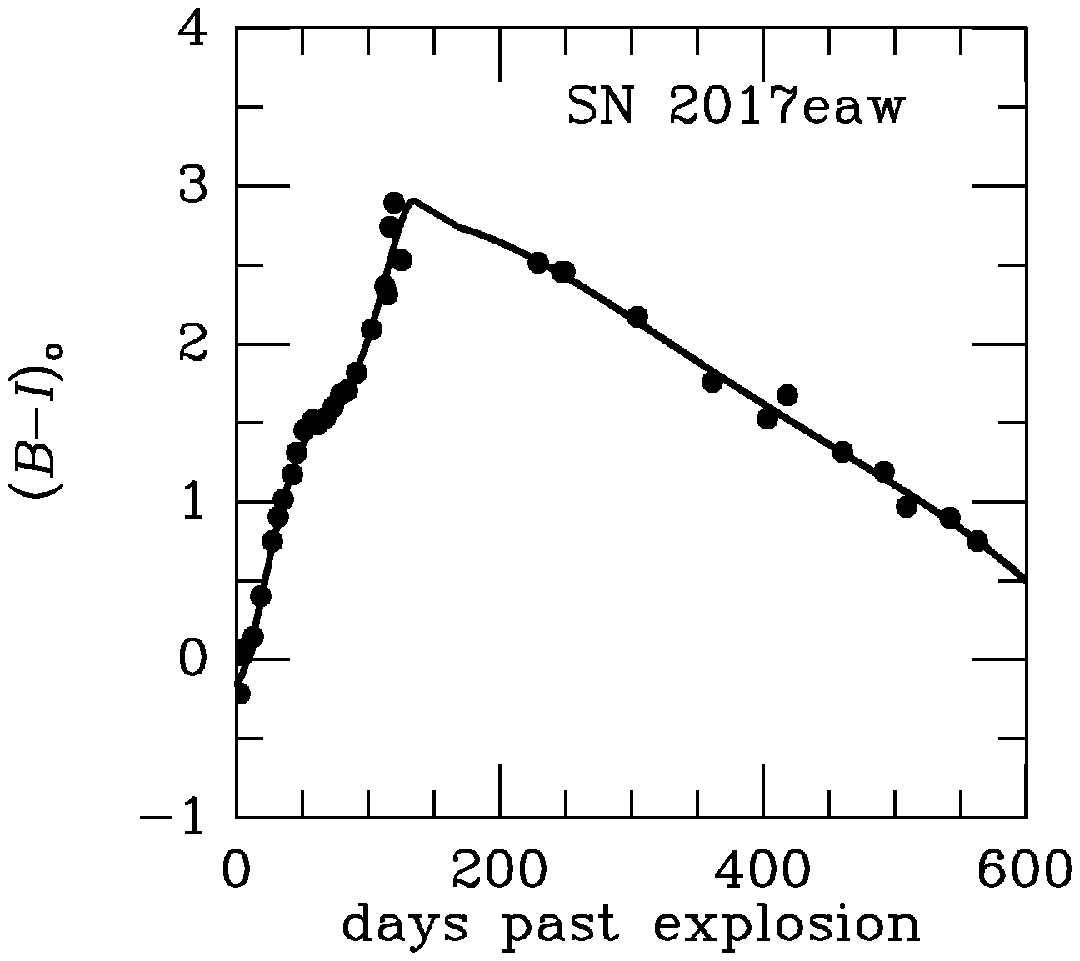} \caption{Polynomial
representation of the $(B-I)_o$ evolution of SN 2017eaw.  Two
polynomials representing times before and after 170 days past the
explosion date are combined and shown with the solid curve. Filled
circles are from Table~\ref{tab:phot}, corrected for an assumed total
reddening of $E(B-V)$ = 0.41 mag.
}
\label{fig:bi0M}
\end{figure}

\noindent
This is based on spectral energy distributions constructed from
available light curves for 21 SNe, and which in general cover the
period from early-on to the end of the plateau phase. The relation is
limited in applicability to the colour range 0.0 $\leq$ $(B-I)_o$
$\leq$ 2.8, which is an issue for SN 2017eaw because of the strong
colour evolution. Even with the high adopted reddening, a few of our
epochs are close to or slightly exceed the upper limit of 2.8
(Figure~\ref{fig:bi0M}). For the purpose of applying equation 4, the
two parts of the $(B-I)_o$ evolution of SN 2017eaw were fitted with
separate 6th or 7th order polynomials and joined at $(B-I)_o$ = 1.70
(solid curve in Figure~\ref{fig:bi0M}), in order to give a smooth
mapping of the evolution to 564 days.

The thick solid curve in Figure~\ref{fig:bolometricM} is the bolometric
light curve of SN 2017eaw based on equation 4 and
Figure~\ref{fig:bi0M}. The bolometric luminosity in ergs s$^{-1}$ is
derived as

$$logL_{bol} = -0.4[B - A_B + {\rm BC}_B - 5logD(cm) + 8.74]\eqno{5}$$

\noindent
where $D$ is the distance in centimeters. The constant is based on an
assumed absolute bolometric magnitude of the Sun of 4.74 (Bessell et
al. 1998). Although equation 4 is technically valid only until the end
of the plateau phase, Figure~\ref{fig:bolometricM} shows how
application of the BC$_B$ polynomial to the full range of SN 2017eaw
epochs (3 to 564 days) yields a bolometric light curve whose shape
strongly resembles that of the quasi-bolometric light curve. As
expected the colour-bolometric light curve is displaced towards higher
values of $L_{bol}$ compared to the quasi-bolometric curve, since it
accounts for missing flux in the UV and IR wavelength domains.

The third approach uses bolometric corrections inferred from the
well-observed supernova 2004et, which occurred in the same galaxy as SN
2017eaw and which we already have shown has $BVRI$ magnitude and colour
evolutions very similar to SN 2017eaw. Maguire et al. (2010) presented
extended observations of SN 2004et, and were able to derive bolometric
corrections, BC($t$), relative to the $V$ and $R$ filters from
$UBVRIJHKL$ observations taken $\approx$5 days to $\approx$120 days
past the explosion. This covers only up to the end of the plateau
phase. Maguire et al. (2010) cover the early tail to $\approx$190 days
using observations of SN 1999em. We use the $R$-band calibration
because it appears to be more homogeneous among different SNe compared
to the $V$-band (Figures 8 and 9 of Maguire et al. 2010). The corrections
were derived as 

$$BC_R(t) = 0.389 + 0.033t -4.27\times 10^{-4} t^2 + 1.80\times 10^{-6}t^3\eqno{6}$$

\noindent
and the resulting bolometric light curve was derived as

$$logL_{bol} = -0.4[R - A_R + {\rm BC}_R(t) - 5logD(cm) + 8.14]\eqno{7}$$

\noindent
This is shown as the short dotted curve in
Figure~\ref{fig:bolometricM}. The constant 8.14 was used by Maguire et
al. in the derivation of the bolometric corrections.

\subsection{Mass of $^{56}$Ni and other parameters}

Having a bolometric light curve allows us to derive important physical
parameters of SN 2017eaw. These can depend on choosing a particular time
of reference, usually something connected with the end of the plateau phase
or the beginning of the radioactive tail phase. These times are shown
in Figure~\ref{fig:VbandM}. 

One set of parameters is described by Nakar et al. (2016); these
include several useful quantities that are concerned with the relative
importance of shock-deposited energy and the decay chain of $^{56}$Ni
on the appearance of the bolometric light curve. The period before the
onset of the radioactive tail is believed to be dominated by
photospheric emission. Nakar et al. note that $^{56}$Ni emission could
play a role on the appearance of the light curve during this phase, by
either extending the plateau or flattening it. If the gamma rays
produced by the $^{56}$Ni decay chain are assumed to be fully confined
and thermalized, then the mass of $^{56}$Ni created in the explosion is
directly proportional to the bolometric luminosity on the tail, which
is assumed to be equal to the energy, $Q_{Ni}$, imparted to the ejected
material by the steps in the $^{56}$Ni decay chain. Nakar et al.'s
equation 1 (see also Sutherland and Wheeler 1984) then gives the
$^{56}$Ni mass as:

$${{M(^{56}{\rm Ni})} \over M_{\odot}} = {L_{bol}(t)\ \ \ {\rm  [tail]} \over {6.45e^{-{t \over 8.8}} + 1.45e^{-{t \over 111.3}}}} \times 10^{-43}\eqno{8}$$

\noindent
where $t$ is the time in days since the explosion. This can be applied at
each epoch along the tail. The standard deviation of these points over
a range of epochs measures how well the decay chain maps to the
bolometric light curve.

\begin{figure}
\includegraphics[width=\columnwidth]{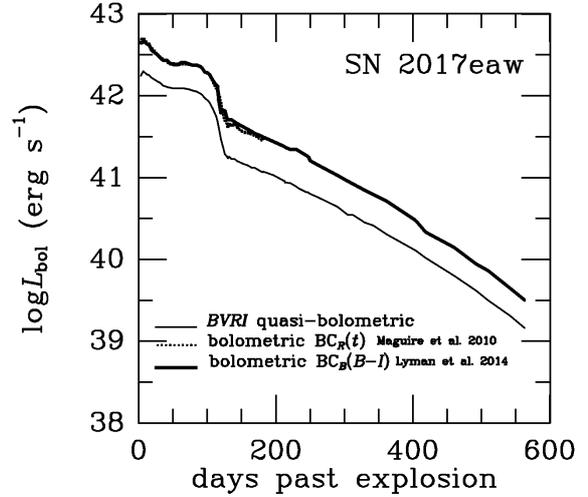}
\caption{The thin solid curve is based on trapezoidal integration of
the interpolated $BVRI$ light curves. The
thick solid curve shows the bolometric light curve based on the
absolute $B$-band light curve and a quadratic representation (Lyman et
al. 2014) of the bolometric correction using the corrected colour index,
$(B-I)_o$, the mapping of which is shown in Figure~\ref{fig:bi0M}. The
dotted curve is the bolometric light curve based on the absolute
$R$-band magnitude and a time-dependent bolometric correction (Maguire
et al. 2010).
}
\label{fig:bolometricM}
\end{figure}

\begin{figure}
\includegraphics[width=\columnwidth]{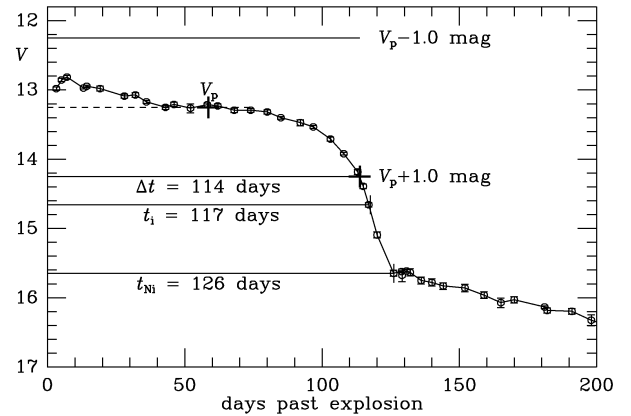}
\caption{Schematic of aspects of the early light curve of a typical Type II-P SN,
showing how certain times are defined (see text).
}
\label{fig:VbandM}
\end{figure}

Table~\ref{tab:nakar} gives the $^{56}$Ni mass implied by equation 8
for the three curves in Figure~\ref{fig:bolometricM}. The
quasi-bolometric light curve gives an underestimate of 0.041$M_{\odot}$
for the $^{56}$Ni mass. The bolometric curve based on equations 4 and 5
gives $M$($^{56}$Ni)=
0.113\rlap{$_{-{0.017}}$}$^{+{0.019}}$$M_{\odot}$, while that based
on equations 6 and 7 gives $M$($^{56}$Ni)=
0.102\rlap{$_{-{0.019}}$}$^{+{0.022}}$$M_{\odot}$.
With these
estimates of the $^{56}$Ni mass, we can derive other parameters
discussed by Nakar et al. (2016) based on time-weighted averages from
$t$ = 0 (explosion date) to $t$ = $t_{\rm Ni}$, the time of the
beginning of the tail phase (Figure~\ref{fig:VbandM}). These parameters
are: $ET$, the time-averaged shock-deposited energy in units of
3$\times$10$^{55}$ erg s; $\eta_{Ni}$, the ratio of the time-weighted
energy injected into the expanding gases due to the $^{56}$Ni decay
chain to $ET$; 2.5log$\Lambda_e$, a parameter telling how fast the
decline and duration of the bolometric light curve would be in the
absence of any $^{56}$Ni emission; and $\Delta M_{25-75}$, a measure of
the decline rate in magnitudes of the bolometric luminosity from day 25
to day 75.  $ET$ is an important parameter that is discussed in more
detail by Shussman et al. (2016), who show that ET $\propto$
$E_{exp}$$^{1\over 2}$ $R_*$ $M_{ej}$$^{1\over 2}$, where $E_{exp}$ is
the explosion energy, $R_*$ is the pre-explosion progenitor radius, and
$M_{ej}$ is the mass of the ejecta.

\begin{table}
\centering
\caption{Nakar et al. (2016) parameters. Col. 1: parameter from Nakar et al. (2016); cols. 2-4: values of parameters for
the curves in Figure~\ref{fig:bolometricM}; col. 5: mean and standard deviation of the same parameters for 13 SNe listed in Table 1 of Nakar et al.
}
\label{tab:nakar}
\begin{tabular}{lrrrr}
\hline
Parameter & Value & Value & Value & mean Nakar\\
bolometric type   &  $BVRI$  & BC$_B$ & BC$_R$ & bolometric\\
1 & 2 & 3 & 4 & 5\\
\hline
Mass Ni$^{56}$/$M_{\odot}$ & 0.041 & 0.113\rlap{$_{-{0.017}}$}$^{+{0.019}}$ & 0.102\rlap{$_{-{0.019}}$}$^{+{0.022}}$ & 0.031$\pm$0.018 \\
$ET$  & 1.25 & 2.27 & 2.43 & 1.03$\pm$0.67 \\
$\eta_{Ni}$ & 0.50 & 0.76 & 0.64 & 0.56$\pm$0.63 \\
2.5log$\Lambda_e$ & 0.73 & 0.91 & 0.87 & 0.79$\pm$0.33 \\
$\Delta M_{25-75}$ & 0.29 & 0.34 & 0.37 & 0.40$\pm$0.38 \\
\hline
\end{tabular}
\end{table}

\begin{table}
\centering
\caption{Litvinova and Nadyozhin (1985) parameters
}
\label{tab:nadyozhin}
\begin{tabular}{lr}
\hline
Parameter & Value \\
1 & 2 \\
\hline
Explosion energy &  15.9\rlap{$_{-{1.1}}$}$^{+{1.3}}$$\times$10$^{50}$ ergs    \\
Mass of ejected envelope &  20\rlap{$_{-{3}}$}$^{+{2}}$$M_{\odot}$ \\
Pre-supernova radius &  536\rlap{$_{-{150}}$}$^{+{200}}$$R_{\odot}$ \\
\hline
\end{tabular}
\end{table}

The values of these parameters derived for SN 2017eaw are listed in
Table~\ref{tab:nakar} and, at least for $\eta_{Ni}$, 2.5log$\Lambda_e$,
and $\Delta M_{25-75}$, are within the ranges found for 13 other SNe by
Nakar et al. (2016, their Table 1), with $ET$ and $\eta_{\rm Ni}$ being
near the higher ends of their ranges. The last column of Table
~\ref{tab:nakar} lists the mean and standard deviation of the Nakar et
al. Table 1 values for comparison. As noted by Nakar et al., these
parameters are less sensitive to the difference between a bolometric
and quasi-bolometric light curve than is the $^{56}$Ni mass. The high
value, $\eta_{\rm Ni}$ = 0.76, obtained for the BC$_B(B-I)$
bolometric light curve implies that $^{56}$Ni contributed 43\% of the
time-weighted integrated luminosity of SN 2017eaw during its
plateau/photospheric emission phase. For the $BC_R(t)$ curve, the
contribution is 39\%. These should be compared with the typical value
of $\approx$30\% for bolometric curves derived by Nakar et al. (2016).
Nakar et al. (2016) also derive these parameters for the
quasi-bolometric light curves of several SNe. For SN 2017eaw we find
$\eta_{\rm Ni}$ = 0.50, which compares very well with the average
value, 0.49, for 5 other supernovae with only $BVRI$ photometry (Nakar
et al., their Table 2). 

The main difference with the Nakar et al. sample is the estimated mass
of $^{56}$Ni. The bolometric curve values in Table~\ref{tab:nakar} are
$\approx$3.5 times higher than the average of the Nakar et al. sample,
and nearly twice the hightest value listed in their Table 1.

Another estimate of the $^{56}$Ni mass for SN 2017eaw can be made based
on the correlation between the absolute $V$-band magnitude {\it on the
plateau} and the $^{56}$Ni mass (Hamuy 2003). Elmhamdi et al. (2003)
discuss the correlation and use 8 SNe to derive

$$log {M(^{56}Ni)\over M_{\odot}} = -0.438 M_V^o(t_i - 35) - 8.46 \eqno{9}$$

\noindent
where $t_i$ is the time when the dropoff rate at the end of the plateau
phase in the $V$-band is maximal (see Figure~\ref{fig:VbandM}), and
$M_V^o(t_i - 35)$ is the corrected absolute magnitude 35 days before
that point in time. From our $V$-band light curve directly, we derive
$t_i$ = 117$\pm$3 days and $M_V^o$ = $-$17.36$\pm$0.04, for which the
above equation gives $M$($^{56}$Ni) =
0.139\rlap{$_{-{0.031}}$}$^{+{0.039}}$$M_{\odot}$, accounting also for
the uncertainty in extinction and distance. This is about 25\% larger
than the $^{56}$Ni mass derived from the bolometric light curves in
Figure~\ref{fig:bolometricM}. Note that Elmhamdi et al. (2003) also
show that the slope, $S$, at $t_i$ also correlates with $log
M(^{56}Ni)$: the larger the value of $S$, the lower the $^{56}$Ni
mass.  Directly from our $V$-band light curve, we get $S$ $\approx$0.14
mag day$^{-1}$, which by equation 2 of Elmhamdi et al. (2003) yields
$M$($^{56}$Ni) = 0.021$M_{\odot}$, considerably less than the
value implied by the $V$-band magnitude on the plateau. Our $V$-band
light curve may not be well-sampled enough to obtain a reliable value
of $S$.\footnote{Van Dyk et al. have better sampling of the rapid
decline phase and obtain $S$=0.089 mag day$^{-1}$, which corresponds to
a $^{56}$Ni mass of 0.04 $M_{\odot}$ according to equation 3 of
Elmhamdi et al. (2003).}

Hamuy (2003) uses the $V$-band luminosity on the tail and a fixed
bolometric correction of 0.26$\pm$0.06 mag to estimate the $^{56}$Ni
mass. Hamuy's equations 1 and 2 applied to our $V$-band light curve from
126 to 290 days gives $M$($^{56}$Ni) = 
0.106\rlap{$_{-{0.022}}$}$^{+{0.027}}$$M_{\odot}$.

Thus, for the distance and reddening we have adopted, and their
uncertainties, the implied $^{56}$Ni mass for SN 2017eaw is
0.115\rlap{$_{-{0.022}}$}$^{+{0.027}}$$M_{\odot}$. This is
$\approx$30\% larger than the amount inferred to have been created in
the SN 1987A explosion (Arnett 1996) but is within the ranges found by
Hamuy (2003) and Elmhamdi et al. (2003). This value is also in good
agreement with Sahu et al.'s (2006) estimate of 0.06$\pm$0.02
$M_{\odot}$ for SN2004et, which, when scaled to our adopted distance,
becomes 0.11$\pm$0.03$M_{\odot}$.

Litvinova and Nadyozhin (1985; see also Nadyozhin 2003) used models of
Type II-P SNe to derive relations connecting the energy of the
explosion, the amount of mass ejected, and the pre-explosion radius of
the star to observable parameters, including the absolute magnitude
$M_{V_p}^o$ in the middle of the plateau, the expansion velocity at
this time, and the time, $\Delta t$, during which the SN was within
$\pm$1.0 mag of $V_p$. The latter parameter is schematically shown in
Figure~\ref{fig:VbandM}. Using Figure 9 of Szalai et al. (2019), we
estimate the expansion velocity for SN 2017eaw to have been 4200 km
s$^{-1}$ at $\approx$60 days past explosion, based on the Fe II 5169
line. The results are summarized in Table~\ref{tab:nadyozhin}. The
explosion energy and pre-supernova radius are found to be
15.9\rlap{$_{-{1.1}}$}$^{+{1.3}}$$\times$10$^{50}$ ergs and
536\rlap{$_{-{150}}$}$^{+{200}}$$R_{\odot}$, respectively. However, the
ejected mass, 20\rlap{$_{-{3}}$}$^{+{2}}$ $M_{\odot}$, greatly exceeds
recent estimates of the total mass of the progenitor of SN 2017eaw
(section 6). Hamuy (2003) discusses the limitations of the Litvinova
and Nadyozhin (1985) formulae, which include assuming no contribution
to the plateau luminosity by $^{56}$Co decay, difficulties in
estimating the photospheric velocity on the plateau, and the assumption
that Type II-P supernovae have blackbody spectra.

\section{The Supernova Progenitor}

SN 2017eaw is one of only a few SNe for which the progenitor has been
identified. Kilpatrick and Foley (2018) describe the detection of the
likely progenitor in {\it Hubble Space Telescope} and {\it Spitzer
Space Telescope} data up to 13 years before the explosion.  These
authors estimate that the progenitor was a red supergiant having an
initial (zero age main sequence) mass of 13\rlap{$_{-2}$}$^{+4}$
$M_{\odot}$. This is based on an assumed distance of 6.72 Mpc, and
would correspond to 14\rlap{$_{-3.5}$}$^{+3}$ $M_{\odot}$ for the
distance, 7.72 Mpc, that we have adopted (Eldridge and Xiao 2019).

Rui et al. (2019) used nine archival {\it HST} images of the site of SN
2017eaw to measure the photometric properties of the progenitor. These
images were obtained at epochs in 2004, 2016, and 2017, with the
majority in 2016. Photometry in different filters was used to determine
the progenitor to be an M4 supergiant with a surface temperature of
3550$\pm$100K and radius of 575$\pm$120$R_{\odot}$. With this and an
estimate of the bolometric luminosity, Rui et al. used an H-R diagram
to obtain a progenitor ZAMS mass of 10-14$M_{\odot}$. This value
assumes a distance of 5.5 Mpc and an extinction $A_V$ = 1.83$\pm$0.59
mag, compared to our adopted values of 7.72 Mpc and 1.27 mag,
respectively. The increase in distance modulus by 0.77 mag is only
partly compensated by the reduction in $A_V$ of 0.56 mag.  If these
latter values were used, the estimated progenitor mass would increase
somewhat. Using pre-explosion HST and Spitzer data and a distance of
7.73 Mpc, Van Dyk et al. (2019) estimated a progenitor mass of
15$M_{\odot}$.

Kilpatrick and Foley (2018) also found evidence for a pre-explosion
circumstellar dust shell about 4000$R_{\odot}$ in radius. Rui et al.
(2019) concluded that the formation of such a shell by mass loss a few
years prior to explosion could have caused an observed reduction in
brightness in 2016. However, Johnson et al. (2018) also examined {\it
HST} images of the progenitor and found little or no evidence for
significant variability within 9 years before the explosion. These
authors also show that the progenitors for three other Type II SNe
similarly show no evidence for significant pre-explosion variability.

\section{Summary}

We have presented $BVRI$ light curves of the classic Type II-P supernova
2017eaw, the tenth supernova discovered in NGC 6946 in 100 years. Our
main findings are:

\noindent
1. The $BVRI$ light and colour curves of SN 2017eaw show all the classic
features of a typical Type II-P supernova.

\noindent
2. These curves strongly resemble those of SN 2004et, another Type II-P
SN that appeared in the same galaxy.

\noindent
3. SN 2017eaw reached an absolute $V$-band magnitude of $-$17.9, based
on an assumed distance of 7.72 Mpc and a reddening of $E(B-V)$ =
0.41$\pm$0.05, and was slightly less luminous than SN 2004et on the
plateau in all four filters.

\noindent
4. The decline in brightness on the early part of the ``tail" is
consistent with ithe luminosity being powered by radioactive decay of
$^{56}$Co into $^{56}$Fe. This is true mainly for the $VRI$ filters,
and less so for the $B$-band.

\noindent
5. The slope of the tail increases past 290 days, possibly indicating the
gradual breakdown of the confinement and thermalization assumption of the
gamma rays released by the radioactivity (Arnett 1996).

\noindent
6. Using several approaches, we have estimated that
0.115\rlap{$_{-{0.022}}$}$^{+{0.027}}$$M_{\odot}$ of $^{56}$Ni was
produced in the explosion of SN 2017eaw, a higher than average
but not extreme value.

\noindent
7. The $^{56}$Ni decay chain contributed 43\% of the time-averaged
bolometric luminosity over the period from 0 to 117 days past the
explosion date (estimated to have occurred on JD 2457886.5). The
remaining 57\% is due to the shock-deposited energy from the explosion
itself that dominates the plateau phase.

\noindent
8. The star that exploded was a red supergiant having an estimated
pre-explosion radius of 536\rlap{$_{-{150}}$}$^{+{200}}$$R_{\odot}$.
The explosion energy was about 1.6$\times$10$^{51}$ ergs. The actual
progenitor has been identified in pre-explosion images.

We thank the referee for helpful comments that improved this paper.
This research has made use of the NASA/IPAC Extragalactic Database
(NED) which is operated by the Jet Propulsion Laboratory, California
Institute of Technology, under contract with the National Aeronautics
and Space Administration.

\section{Appendix}

Because NGC 6946 is such a prolific producer of SNe, several published
sources having photometry of local field stars are available for
comparison with our Table 2 system. One source, Buta (1982), has only
$UBV$ photometry, but Table 1 of that paper includes 10 stars in common
with Table~\ref{tab:stds}. Botticella et al. (2009) present $UBVRI$
photometry of 26 numbered stars they used for photometry of SN 2008S,
of which 5 are included in Table~\ref{tab:stds}. These authors observed
the same local standards as did Pozzo et al. (2006), but added $U$ and
$B$. Sahu et al. (2006) based their photometry of SN 2004et on 8 local
standards, only 1 of which overlaps Table~\ref{tab:stds}. In order to
include the Sahu et al. photometry in our comparison,
Table~\ref{tab:sahustds} lists the magnitudes and colours of the eight
Sahu et al. local standards using the same images and the same
transformation and extinction coefficients as were used for the stars
in Table~\ref{tab:stds}. All are faint standards and Sahu stars 6-8
have an especially low signal-to-noise compared to most of the stars in
Table~\ref{tab:stds}. Our comparison is based on Sahu et al. stars
1-5.

The results are summarized in Table~\ref{tab:compstds}. In general, we
find that the mean differences between the other sources and the
photometric system represented by Table~\ref{tab:stds} are generally
less than $\pm$0.04 mag. Because NGC 6946 is likely to host many more
SNe, we anticipate improving the Table~\ref{tab:stds} photometric system 
in future studies.

\begin{table}
\centering
\caption{Photometry of local standards used by Sahu et al.
(2006)
}
\label{tab:sahustds}
\begin{tabular}{lrrrr}
\hline
Sahu et al. & $V$ & $B-V$ & $V-R$ & $R-I$  \\
No.     & m.e. & m.e. & m.e.  & m.e.  \\
 1 & 2 & 3 & 4 & 5 \\
\hline
        1 &   15.185 &    0.731    &  0.501  &   0.433  \\
          &    0.044 &    0.108    &  0.061  &   0.028  \\
        2 &   13.772 &    0.660    &  0.430  &   0.411  \\
          &    0.015 &    0.040    &  0.014  &   0.012  \\
        3 &   14.262 &    1.367    &  0.716  &   0.665  \\
          &    0.020 &    0.109    &  0.028  &   0.013  \\
        4 &   14.740 &    0.728    &  0.500  &   0.479  \\
          &    0.017 &    0.053    &  0.012  &   0.020  \\
        5 &   14.833 &    0.793    &  0.483  &   0.492  \\
          &    0.028 &    0.051    &  0.028  &   0.021  \\
        6 &   16.118 &    0.978    &  0.708  &   0.563  \\
          &    0.156 &    0.289    &  0.131  &   0.037  \\
        7 &   16.401 &    0.604    &  0.537  &   0.513  \\
          &    0.067 &    0.101    &  0.082  &   0.025  \\
        8 &   16.400 &    0.468    &  0.549  &   0.391  \\
          &    0.074 &    0.246    &  0.100  &   0.040  \\
\hline
\end{tabular}
\end{table}

\begin{table}
\centering
\caption{Comparisons of Table~\ref{tab:stds} system magnitudes with published
local photometry from Buta (1982), Sahu et al. 2006, and Botticella et al. (2009).
The sense is $\Delta$ = (published values) $-$ (Table~\ref{tab:stds} system values).
}
\label{tab:compstds}
\begin{tabular}{lrrr}
\hline
Filter/colour       & $<\Delta>$ & $<\Delta>$ & $<\Delta>$  \\
       & Buta 1982  & Sahu et al. & Botticella et al. \\
 1     & 2            & 3        & 4      \\
\hline
  $B$ &  0.018$\pm$0.009 & $-$0.036$\pm$0.032 & $-$0.037$\pm$0.041 \\
  $V$ & $-$0.002$\pm$0.005 & $-$0.021$\pm$0.011 & $-$0.019$\pm$0.019 \\
  $R$ & .................. &  0.004$\pm$0.007 &  0.020$\pm$0.006 \\
  $I$ & .................. &  0.006$\pm$0.012 &  0.037$\pm$0.008 \\
$B-V$ &  0.020$\pm$0.007 & $-$0.016$\pm$0.032 & $-$0.019$\pm$0.031 \\
$V-R$ & .................. & $-$0.025$\pm$0.005 & $-$0.039$\pm$0.021 \\
$R-I$ & .................. & $-$0.002$\pm$0.008 & $-$0.017$\pm$0.011 \\
$n$   & 10                 & 5                  &            5       \\
\hline
\end{tabular}
\end{table}

\end{document}